\documentclass[iop,appendixfloats,numberedappendix]{emulateapj-rtx4}

\usepackage{subfigure}
\usepackage{floatflt,graphicx}
\usepackage{amsmath}
\usepackage{array}
\usepackage{verbatim}
\usepackage{rotating}
\usepackage[section] {placeins}

\RequirePackage{epsf,graphicx}

\newcommand{\HI}{H\,{\sc i}}
\newcommand{\HII}{H\,{\sc ii}}

\newcommand{\arcs}{\arcsec }

\shorttitle{Galaxy And Mass Assembly (GAMA): Mid-Infrared Properties}
\shortauthors{Cluver et al.}

\begin{document}

\title{Galaxy And Mass Assembly (GAMA): Mid-Infrared Properties and Empirical Relations from {\it WISE}}

\author{M.E. Cluver\altaffilmark{1,2},\email{mcluver@ast.uct.ac.za} T.H. Jarrett\altaffilmark{1}, A.M. Hopkins\altaffilmark{3}, S.P. Driver\altaffilmark{4,5}, J. Liske\altaffilmark{6}, M.L.P. Gunawardhana\altaffilmark{7,8,3}, E.N. Taylor\altaffilmark{9}, A.S.G. Robotham\altaffilmark{4}, M. Alpaslan\altaffilmark{5}, I. Baldry\altaffilmark{10}, M.J.I. Brown\altaffilmark{11}, J.A. Peacock\altaffilmark{12}, C.C. Popescu\altaffilmark{13}, R.J. Tuffs\altaffilmark{14},  A.E. Bauer\altaffilmark{2},   J. Bland-Hawthorn\altaffilmark{8},  M. Colless\altaffilmark{15}, B. Holwerda\altaffilmark{16}, M.A. Lara-L{\'o}pez\altaffilmark{2}, K. Leschinski\altaffilmark{17}, A.R. L{\'o}pez-S{\'a}nchez\altaffilmark{3,18}, P. Norberg\altaffilmark{7}, M. Owers\altaffilmark{2}, L. Wang\altaffilmark{7}, S.M. Wilkins\altaffilmark{20}}

\slugcomment{Accepted to ApJ: 4 January 2014}

\altaffiltext{1}{Department of Astronomy, University of Cape Town, Private Bag X3, Rondebosch, 7701, South Africa}
\altaffiltext{2}{ARC Super Science Fellow, Australian Astronomical Observatory, PO Box 915, North Ryde, NSW 1670, Australia}
\altaffiltext{3}{Australian Astronomical Observatory, PO Box 915, North Ryde, NSW 1670, Australia}
\altaffiltext{4}{International Centre for Radio Astronomy Research (ICRAR), University of Western Australia, 35 Stirling Highway, Crawley, WA 6009, Australia}
\altaffiltext{5}{SUPA, School of Physics and Astronomy, University of St Andrews, North Haugh, St Andrews, Fife, KY16 9SS, UK}
\altaffiltext{6}{European Southern Observatory, Karl-Schwarzschild-Str. 2, 85748 Garching, Germany}
\altaffiltext{7}{Institute for Computational Cosmology, Department of Physics, Durham University, South Road, Durham DH1 3 LE, U.K.}
\altaffiltext{8}{Sydney Institute for Astronomy (SIfA), School of Physics, University of Sydney, NSW 2006, Australia}
\altaffiltext{9}{School of Physics, the University of Melbourne, Parkville, VIC 3010, Australia}
\altaffiltext{10}{Astrophysics Research Institute, Liverpool John Moores University, IC2, Liverpool Science Park, 146 Brownlow Hill, Liverpool, L3 5RF}
\altaffiltext{11}{School of Physics, Monash University, Clayton, Victoria 3800, Australia}
\altaffiltext{12}{Institute for Astronomy, University of Edinburgh, Royal Observatory, Edinburgh EH9 3HJ, UK}
\altaffiltext{13}{Jeremiah Horrocks Institute, University of Central Lancashire, PR1 2HE, Preston, UK}
\altaffiltext{14}{Max Planck Institut fuer Kernphysik, Saupfercheckweg 1, 69117 Heidelberg, Germany}
\altaffiltext{15}{The Australian National University, Canberra, ACT 2611, Australia}
\altaffiltext{16}{University of Leiden, Sterrenwacht Leiden, Niels Bohrweg 2, NL-2333 CA Leiden, The Netherlands}
\altaffiltext{17}{Institute for Astrophysics, University of Vienna, T{\" u}rkenschanzstra{\ss}e 17, A-1180 Vienna}
\altaffiltext{18}{Department of Physics and Astronomy, Macquarie University, NSW 2109, Australia}
\altaffiltext{19}{Astronomy Centre, Department of Physics and Astronomy, University of Sussex, Brighton, BN1 9QH, U.K.}

\begin{abstract}

The Galaxy And Mass Assembly (GAMA) survey furnishes a deep redshift catalog that, when combined with the Wide-field Infrared Explorer ({\it WISE}), allows us to explore for the first time the mid-infrared properties of $> 110, 000$ galaxies over 120\,deg$^2$ to $z\simeq0.5$. In this paper we detail the procedure for producing the matched GAMA-{\it WISE} catalog for the G12 and G15 fields, in particular characterising and measuring resolved sources; the complete catalogs for all three GAMA equatorial fields will be made available through the GAMA public releases.
The wealth of multiwavelength photometry and optical spectroscopy allows us to explore empirical relations between optically determined stellar mass (derived from synthetic stellar population models) and 3.4\micron\ and 4.6\micron\ {\it WISE} measurements. Similarly dust-corrected H$\alpha$-derived star formation rates can be compared to 12\micron\ and 22\micron\ luminosities to quantify correlations that can be applied to large samples to $z<0.5$. To illustrate the applications of these relations, we use the 12\micron\ star formation prescription to investigate the behaviour of specific star formation within the GAMA-{\it WISE} sample and underscore the ability of {\it WISE} to detect star-forming systems at $z\sim 0.5$. Within galaxy groups (determined by a sophisticated friends-of-friends scheme), results suggest that galaxies with a neighbor within $100\,h^{-1}$kpc have, on average, lower specific star formation rates than typical GAMA galaxies with the same stellar mass.

\end{abstract}


\section{Introduction}

Spanning almost six decades and counting, large area surveys have revolutionised our view of the structure and evolution of the universe, for example, the Palomar Observatory Sky Survey (POSS II, Reid et al. 1991), the IIIa-J SRC Southern Sky Survey, the Infrared Astronomy Satellite (IRAS) Sky Survey Atlas (Wheelock et al. 1994), 2dF Galaxy Redshift Survey \citep[2dFGRS,][]{Coll01}, the Sloan Digital Sky Survey (SDSS, Abazajian et al. 2003), the \HI\ Parkes All Sky Survey (HIPASS, Meyer et al. 2004), the 6dF Galaxy Survey \citep[6dFGS,][]{Jon04}, the 2 Micron All Sky Survey (2MASS, Skrutskie et al. 2006) and the UKIRT Infrared Deep Sky Survey \citep[UKIDSS,][]{Law07}. Uncovering the properties of large galaxy populations has proved essential to our understanding of how galaxies develop and transform. Increasing sensitivity and scale in the time domain, as with Skymapper and the planned LSST (Large Synoptic Survey Telescope) holds the promise of enormous future scientific returns. In the near future, leading-edge radio telescopes and planned strategic surveys will add a key ingredient to the mix -- \HI\ and continuum measurements will trace neutral gas reservoirs and activity to unprecedented depth, sky coverage and resolution with WALLABY and EMU on ASKAP (Australian Square Kilometer Array Pathfinder), Apertif-WNSHS, LADUMA \citep{Hol10} on MeerKAT, and the JVLA.

In order to better understand the key physics at work in star-forming galaxies, and crucially the efficacy of different physical mechanisms, one requires the combination of extended area and sufficient depth in order to capture a cosmologically and evolutionarily significant volume. The GAMA \citep[Galaxy and Mass Assembly,][]{Driv09, Driv11} survey provides exactly this laboratory. At its heart, GAMA is an optical spectroscopic survey of up to $\sim$ 300 000 galaxies \citep{Hop13} obtained at the Anglo-Australian Telescope situated at Siding Spring Observatory. Three equatorial fields (G09, G12 and G15) covering 180\,deg$^2$ sample large-scale structure to a redshift of z$\simeq 0.5$, with overall median redshift of z$\simeq 0.3$, and two southern fields (G02 and G23). Multiwavelength ancillary data from ultraviolet to far-infrared wavelengths provides comprehensive SED (Spectral Energy Distribution) wavelength real estate.
At mid-infrared wavelengths it falls to {\it WISE}, the Wide-field Infrared Survey Explorer, to span the near-infrared transition from stellar to dust emission.

Launched in December 2009, {\it WISE} surveyed the entire sky in four mid-infrared bands: 3.4\micron\, 4.6\micron, 12\micron\ and 22\micron\ \citep{Wr10}. In the local Universe, the 3.4\micron\ (W1) and 4.6\micron\ (W2) bands chiefly trace the continuum emission from evolved stars \citep[see, for example,][]{Meidt12}. The W1 band is the most sensitive to stellar light, typically reaching $L^{\ast}$ depths to $z\simeq 0.5$. 
The W2 band is additionally sensitive to hot dust; hence,  this makes the 3.4\micron\ $-$ 4.6\micron\ color very sensitive to galaxies dominated globally by active galactic nucleus (AGN) emission \citep[see e.g.,][]{Jar11, Ster12}.
The 12\micron\ (W3) band is broad \citep[see][for relative system response curves]{Wr10} and can trace both the 9.7\micron\ silicate absorption feature, as well as the 11.3\micron\ PAH (polycyclic aromatic hydrocarbon) and $[$Ne{\sc ii}$]$ emission lines. Lastly, the W4 band traces the warm dust continuum at 22\micron, sensitive to reprocessed radiation from star formation and AGN activity.  {\it WISE} is thus optimally suited to study the diverse emission mechanisms of galaxies.

{\it WISE} is confusion-noise-limited --  the structure of the background (e.g. distant galaxies and scattered light) increases noise by contributing flux -- and under these conditions point spread function (PSF) profile-fitted photometry performs robustly \citep
{Mar12}. For unresolved sources, this is the adopted method of obtaining photometry and accordingly the {\it WISE} Source Catalog is optimised and calibrated for point sources. The angular resolution of {\it WISE} frames, used for determining the profile photometry measurement (keyword {\it w$\star$mpro} in the {\it WISE} All-Sky Data Release), is 6.1\arcs, 6.4\arcs, 6.5\arcs and 12.0\arcs\ for W1, W2, W3 and W4, respectively.   The {\it WISE} public-release data products also includes an image Atlas of tiled mosaics; however, the angular resolution is slightly degraded in the Atlas images because they are convolved with a matched filter to optimize point-source detection. The crucial point here is that the {\it WISE} catalogs and images are not well-suited to characterization of resolved sources.  Resolved sources will either be detected and measured as conglomerations of several point sources, or have their flux grossly underestimated by the PSF profile.   As a consequence, it is left to the community to properly measure sources that are resolved by {\it WISE} (i.e., nearby galaxies).  In this work we have produced new, better-optimized images to extract resolved galaxy measurements from the GAMA fields to complement the point-source measurements obtained from the public-release catalogs. 

A detailed {\it WISE} study of several nearby ($<60$ Mpc) galaxies \citep{Jar13} has highlighted how {\it WISE} data can be used in multiwavelength studies of star formation and interstellar medium (ISM) conditions, as well as tracing global properties such as stellar mass and star formation. The work of \citet{Don12} showed the power of combining {\it WISE} with a large-area survey (SDSS); they investigated the effect of star formation- and AGN-activity on the {\it WISE} properties of $>95, 000$ galaxies, including calibrations using the 12\micron\ and 22\micron\ luminosities. These empirical calibrations have also been investigated in 22\micron-selected galaxies in SDSS galaxies with $z<0.3$ by \citet{Lee13}.
{\it WISE} colours have proved to be an excellent AGN selection tool \citep{Jar11, Ster12, Ass13}, and as a diagnostic for determining the accretion modes amongst radio-loud AGN \citep{Gur13}.  \citet{Yan13} combine {\it WISE} and SDSS to provide a phenomenological characterization for {\it WISE} extragalactic sources.

In this paper we harness the power of the GAMA survey, and its value-added data products, cross matching GAMA galaxies in two completed GAMA fields (the G12 and G15 regions) to their {\it WISE} counterparts. The G09 region is not considered in this analysis, but will be included as part of the GAMA-{\it WISE} data release.  

GAMA counterparts at low redshift will be predominantly resolved, particularly in the W1 and W2 bands. Unlike previous studies, we include robust measurements of resolved sources in these fields and distinguish between resolved and unresolved systems to determine the photometry most appropriate for a given source. Since GAMA is highly spectroscopically complete to $r_{\rm AB}<19.8$ within the two GAMA regions, we are able to push {\it WISE} to higher redshifts than previously possible for wide-field surveys. We explore empirical relationships for stellar mass and star formation rates, the color distribution of {\it WISE} sources in GAMA and their behaviour in dense environments. The paper is arranged as follows: in \S\,2 we describe the data used in this analysis, \S\,3 discusses how the GAMA-{\it WISE} matched catalog is constructed, \S\,4 contains results derived from the combined surveys and \S\,5 illustrates scientific applications of the catalog. Conclusions are presented in \S\,6.

The cosmology adopted throughout this paper is $H_0 = 70 $\,km\,s$^{-1}$ Mpc$^{-1}$, $h={\rm H}_0/100$, $\Omega_M=0.3$ and $\Omega_{\Lambda} = 0.7$. All magnitudes are in the Vega system ({\it WISE} calibration described in Jarrett et al. 2011) unless indicated explicitly by an AB subscript.

\section{Data}

\subsection{GAMA}

Our sample is drawn from the G12 and G15 equatorial fields of the GAMA II survey \citep[][]{Driv09,Driv11} combining high spectroscopic completeness ($\simeq$ 97\%) to a limiting magnitude of $r_{\rm AB}=19.8$, with a wealth of ancillary photometric data. Details of target selection for the survey are outlined in \citet{Bal10} and optimal tiling for the survey in \citet{Robot10}. Spectra were obtained primarily with the 2dF instrument mounted on the 3.9-m Anglo-Australian Telescope, and additionally from the Sloan Digital Sky Survey \citep[SDSS DR7; ][]{Ab09}. Reduction and analysis of the spectra is discussed in detail in \citet{Hop13} and Liske et al. (in prep.). 

Photometric data for galaxies within the G12 and G15 volumes (ApMatchedCatv05) is drawn from SDSS imaging  ($u,g,r,i,z$) as outlined in \citet{Hill11} and VISTA VIKING ($ZYJHK$) as detailed in Driver et al. (in prep.). Photometry is corrected for Galactic foreground dust extinction using \citet{Sch98}. Stellar mass estimates (StellarMassesv15) are derived from matched-aperture photometry using synthetic stellar population models from \citet{BC03} as detailed in \citet{Tay11}. We take advantage of the completed observations in the equatorial fields and updated redshifts (Baldry et al., in prep.) and stellar masses. Throughout our analysis we only use sources with reliable redshifts i.e. $nQ \ge 3$ \citep{Driv11}.

Star formation rates (SFRs) can be derived from the H$\alpha$ equivalent width, combined with the $r$-band absolute magnitude, and are available for the GAMA phase-I survey ($r_{\rm AB}<19.4$ in G15 and $r_{\rm AB}<19.8$ in G12) as determined in \citet{Gun11, Gun13}. Corrections are applied for the underlying Balmer stellar absorption, dust obscuration and fibre aperture effects. To maintain consistency we only use SFRs from \citet{Gun13} derived for galaxies with redshifts matching in both GAMA I and GAMA II ($\simeq 10\%$ were excluded because of this criterion).

\subsection{{\it WISE} Image Construction}

The {\it WISE} mission provides `Atlas' Images via its data repository, which are co-added and interpolated from the multiple single exposure frames taken during the survey \citep{Cut12}. These 1.56$\arcdeg \times 1.56\arcdeg$ mosaics have a 1.375\arcs\ pixel scale, but the resampling and co-addition method is optimised for point source detection; Atlas images have beam sizes of 8.1\arcs, 8.8\arcs, 11.0\arcs\ and 17.5\arcs\ for the W1, W2, W3 and W4 bands, respectively. To better preserve the native resolution of the single frames we employ a mosaic construction that uses the `drizzle' resampling technique.
Variable-Pixel Linear Reconstruction or drizzling co-addition algorithm using a tophat PRF (point response function) can be used to improve the spatial resolution compared to the nominal Atlas Images \citep{Jar12}. Drizzled image mosaics, 1.3$\times$1.3 degrees in size, were created using the software package ICORE \citep{Mas13} achieving a resolution of 5.9\arcs, 6.5\arcs, 7.0\arcs\ and 12.4\arcs\ in the 3.4\micron, 4.6\micron, 12\micron\ and 22\micron, respectively. Within ICORE, background level offset-matching, flagging and outlier rejection and co-addition using overlap area-weighted interpolation ensures optimal background stability. The number of images that are combined is dependent on the depth of coverage and additional orbits, a function of the field's location relative to the ecliptic. For the GAMA fields the coverage is typically 12 and 24 frames for the G12 and G15 fields, respectively, where G15 benefits from additional scans in that region arising from multiple epoch {\it WISE} orbits. 

\begin{figure*}[!thb]
\begin{center}
\includegraphics[width=15cm]{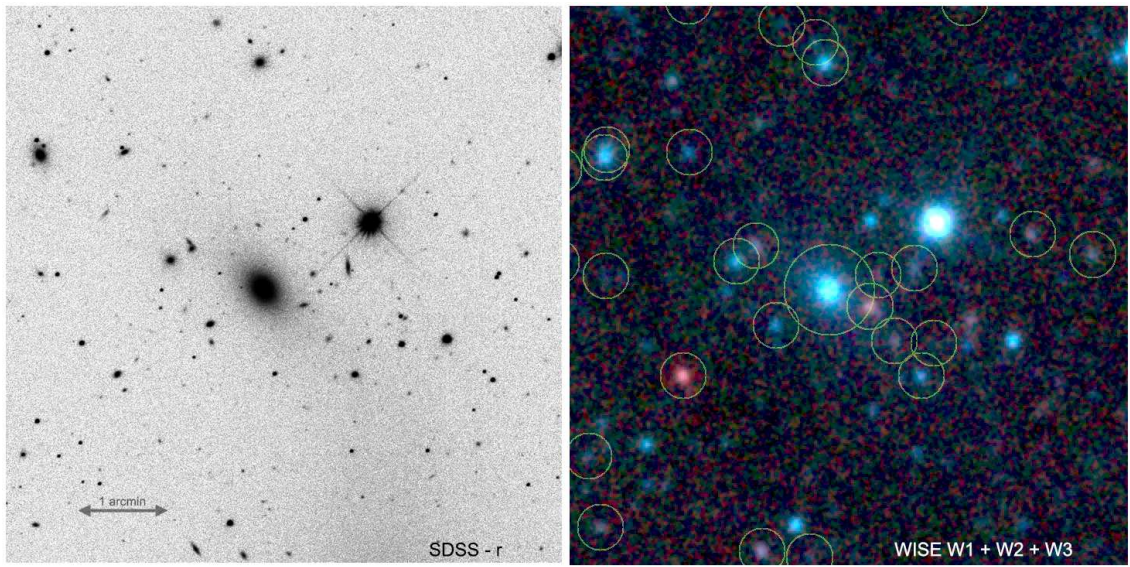}

\caption{A representative 6\arcmin\ $\times$ 6\arcmin\ cutout from the G15 field with the SDSS $r$-band image (left) and the three-color (3.4\micron, 4.6\micron, 12\micron) {\it WISE} image of the same region (right). WISE-GAMA galaxies are circled, demonstrating a mix of resolved and point-like sources. }
\label{fig1}
\end{center}
\end{figure*}

\subsection{Cross Match to GAMA G12 and G15}

Galaxies were extracted from the G12 and G15 GAMA II catalogs of observed sources (i.e. TilingCatv41 with redshifts as in SpecAllv21 and $z>0.001$) -- 69,693 and 64,822 for G12 and G15, respectively, and cross-matched to the {\it WISE} All-Sky Catalog using the NASA/IPAC Infrared Science Archive (IRSA) and a 3\arcs\ cone search radius. The number of unique objects in each field were 60,645 GAMA matches for G12 and 58,199 for G15, i.e., a 87.0\% and 89.8\% match rate for G12 and G15, respectively, with a higher match rate in G15  due to the increased {\it WISE} depth (from additional scans) and therefore sensitivity. Further details are provided in Table \ref{match} and a region of G15 shown in Figure \ref{fig1} to illustrate the difference between the optical and mid-infrared sky.
 
For this analysis, we rely on several parameters output by the {\it WISE} All-Sky Catalog; these are listed in Table \ref{param} with a brief explanation.

\begin{table}[!htb]
\caption{{\it WISE} Cross-Match Statistics for each Band \label{match}}
\begin{center}
\begin{tabular}{l l l l l}
\hline
\\[0.25pt]
\multicolumn{5}{c}{G12 (total: 60,645 sources)}\\
Band & Detections  & SNR$^{1}$ $>2$  & SNR$ >5$  & Upper Limits$^{\dagger}$ \\
 W1  &  100\%   & 100\%  & 99.9\%   & 0\%\\
 W2  &  100\%   & 96\%   & 77\%   &  4\% \\
 W3  &  99\%    &  47\%  &  17\%   &   52\%\\
 W4  &   98\%   &  19\%  & 2\%    &    79\% \\
 \\[1pt]
 \hline
 \\
 \multicolumn{5}{c}{G15 (total: 58,199 sources)}\\
Band & Detections   & SNR$>2$   &SNR$ >5$  & Upper Limits$^{\dagger}$ \\
 W1  &  100\%  &  100\%  &   99.9\%   & 0.01\%\\
 W2  &  100\%   &  98\%   &  88\%   &  2\% \\
 W3  &   99\%  &   58\%   &  29\%   &   42\%\\
 W4  &   98\%   &  19\%   &   3\%    &    78\% \\
\\[1pt]
\hline
\\
\multicolumn{5}{l}{$^{1}$ Signal-to-noise ratio as measured by {\it w$\star$mpro}} \\
\multicolumn{5}{l}{$^{\dagger}$ Detections with 2$\sigma$ upper limits}\\

\end{tabular}
\end{center}
\end{table}

\begin{table}[!htb]

\caption{Parameters from the {\it WISE} All-Sky Catalog  \label{param}}
\begin{center}
\begin{tabular}{l l}
\hline
\\[0.25pt]
{\it designation}
& unique {\it WISE} source designation\\
 {\it ra} (deg) &
 Right Ascension (J2000)\\
 {\it dec} (deg) &
Declination (J2000)\\
{\it sigra} (arcsec) &
 uncertainty in RA
\\ {\it sigdec} (arcsec) 
& uncertainty in DEC
\\ {\it w$\star$mpro} (mag) 
& instrumental profile-fit photometry magnitude
\\ {\it w$\star$sigmpro} (mag) 
& instrumental profile-fit photometry uncertainty 
\\ {\it w$\star$rchi2}
& instrumental profile-fit photometry reduced $\chi^2$

\\ {\it nb}
& number of blend components used in each fit
\\ {\it na}
& active deblend flag (=1 if actively deblended)
\\ {\it xscprox} (arcsec) 
& distance between source center and XSC$^{\dagger}$ galaxy
\\ {\it w$\star$rsemi}
& (scaled) semi-major axis of galaxy from XSC$^{\dagger}$
\\ {\it w$\star$gmag}
& elliptical aperture mag of extracted galaxy
\\[1pt]
\hline
\\
\multicolumn{2}{l}{$\star$= 1,2,3,4}\\
\multicolumn{2}{l}{$^{\dagger}$: 2MASS Extended Source Catalog \citep{Jar00}}\\
\end{tabular}
\end{center}
\end{table}
 
\section{Sources Resolved by {\it WISE}}

A key feature of a catalog containing {\it WISE} photometry is determining which of the galaxies are resolved and then ensuring their fluxes are measured robustly. A broad indication is given by the reduced $\chi^2$ ({\it w$\star$rchi2}) of the profile-fit solution, where high values of the {\it w$\star$rchi2} parameter indicate that the wpro algorithm measurement is underestimating the flux of the source. Sources with {\it w$\star$rchi2} $>2$ are often resolved, with {\it w$\star$rchi2} $>5$ usually indicating well-resolved systems.

Unfortunately {\it w$\star$rchi} is not by itself a robust measure of `resolvedness', particularly where sources have a low signal to noise.  The `reduced' metric tends to unity as noise begins to dominate the measurement.
Moreover, the {\it w$\star$mpro} fitting process can be fooled for sources with relatively bright cores that are just being resolved by {\it WISE} -- this happens with 2MASS compact extended sources which have low {\it w$\star$rchi2}, but the {\it w$\star$mpro} and isophotal photometry can be systematically offset.

Even though 2MASS has superior (2$\times$) spatial resolution compared to {\it WISE}, with its greater sensitivity {\it WISE} is able to resolve many nearby, relatively small galaxies. Therefore, all 2MASS Extended Source Catalog \citep[XSC,][]{Jar00} sources should be tested to determine if they are resolved by {\it WISE} (see \S\,\ref{rfuzzy}).
For galaxies not in the 2MASS XSC, values of {\it w$\star$rchi2} $\ge 2$ should be used as an initial selection for resolved objects. Caution is required, however, since blended objects will also satisfy this criterion and can be a source of false positives, notably when the stellar confusion is significant.

\subsection{Source Characterization of Potentially Resolved Sources}

Sources are measured using custom software adapting tools and algorithms developed for the 2MASS XSC \citep{Jar00} and {\it WISE} photometry pipelines \citep{Jar11, Cut12, Jar13}. The process is semi-automated in that photometry measurements are automated, but each result is assessed by visual inspection with intervention where necessary.

The first step is to remove point sources by PSF subtraction which preserves the structure of the background. If necessary surrounding contaminating sources are masked (e.g., bright stars). The local background is estimated from pixel value (trimmed average) distribution that lies within an elliptical annulus located just outside of an `active region' which represents the image area that contains measurable light from the galaxy.  The active region is not initially known, but it determined through successive iteration of the characterisation process until convergence is reached.

The galaxy is modeled using an ellipsoid built from azimuthally averaging the (background-subtracted) surface brightness; any masked pixels are recovered using a weighted combination of the local background (to the masked source) and the galaxy model. The best fit axis ratio and ellipticity are determined using the 3$\sigma$ isophote and the galaxy shape is defined by this isophote and is assumed to be fixed at all radii.  The primary `isophotal' photometry (W$\star$iso) is then measured from the 1$\sigma$ (of the background RMS) elliptical isophote, capturing over 90\% of the total light (see Jarrett et al 2013).  Other measurements include a double S\'{e}rsic fit, to the inner galaxy region (i.e., the bulge) and the outer region (i.e., the disk), thus allowing estimation of the total integrated flux that extends beyond the 1$\sigma$ isophote.  
 Since {\it WISE} is confusion-noise limited, we track the photometry using a curve-of-growth table. Where a large mismatch occurs between the isophotal radius and the radius where the change in flux is less than 2\% (the `convergence point'), the size of the active region is automatically decreased. This usually occurs where background levels are elevated due to a neighboring bright source which contaminates the background level.  The process is iterated to adjust the active region until the source measurements have converged.

Once all the sources in a field are measured and modeled accordingly, the process is rerun to effectively perform galaxy-galaxy deblending. Where a measurement exists for a galaxy when measuring an adjacent source, the galaxy model is subtracted in addition to neighboring point sources. This mitigates contamination from nearby resolved sources.


\subsection{The Resolved Parameter {\it Rfuzzy}}\label{rfuzzy}

The {\it WISE} bands at 3.4\micron\ and 4.6\micron\ achieve the best spatial resolution, at 6.1\arcs\ and 6.4\arcs, respectively, and capture extended light arising from evolved stars in the bulge and disk regions. Since {\it w1rchi2} cannot discriminate well between faint, resolved and unresolved systems, we use the 2nd order intensity-weighted moment to describe the shape (major and minor axis) of the object and measure the intensity-weighted moment of the object on either side of the main axis. 

If the object is symmetric and resolved on both sides of the major axis (as with a nominal resolved source), then each moment will have a resolved signature (i.e., the moment is large compared to a point source). If the object is, for example, a star blended with a galaxy, then their moments will be different (one half will have a moment that is point-like, the other resolved). Thus, taking the minimum moment value between the two halves, determines whether the source is resolved. Fuzzy objects will have a large moment regardless of which major axis half that is measured; stars or double stars will have a minimum moment that appears to be unresolved.
The performance of the Rfuzzy parameter is illustrated in \S\,\ref{rfuzz_sec} of the appendix.

\subsection{Catalog of Resolved Sources}

\begin{figure*}[!th]
\begin{center}
\subfigure[3.4\micron]{\includegraphics[width=8cm]{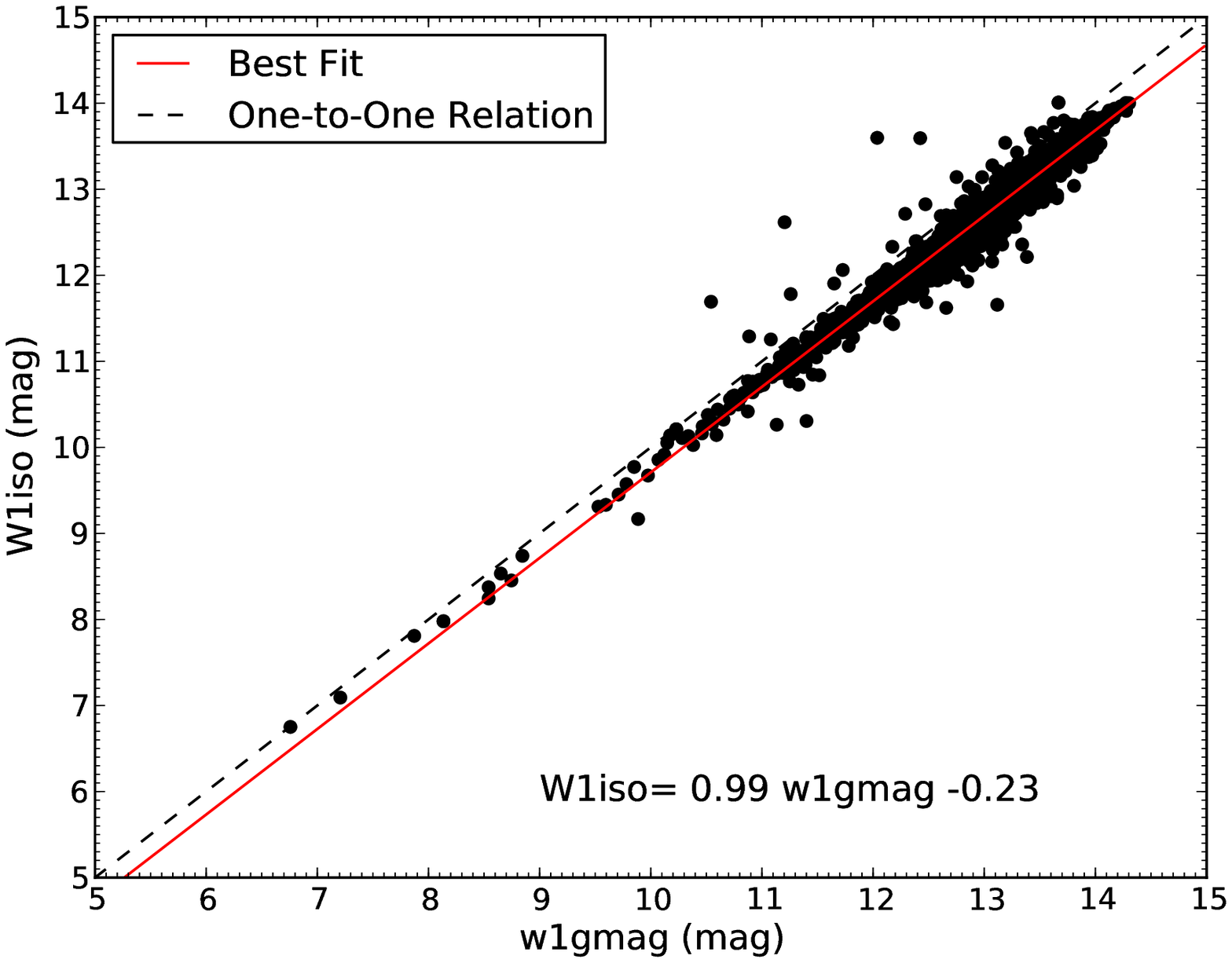}}
\hfill
\subfigure[4.6\micron]{\includegraphics[width=8cm]{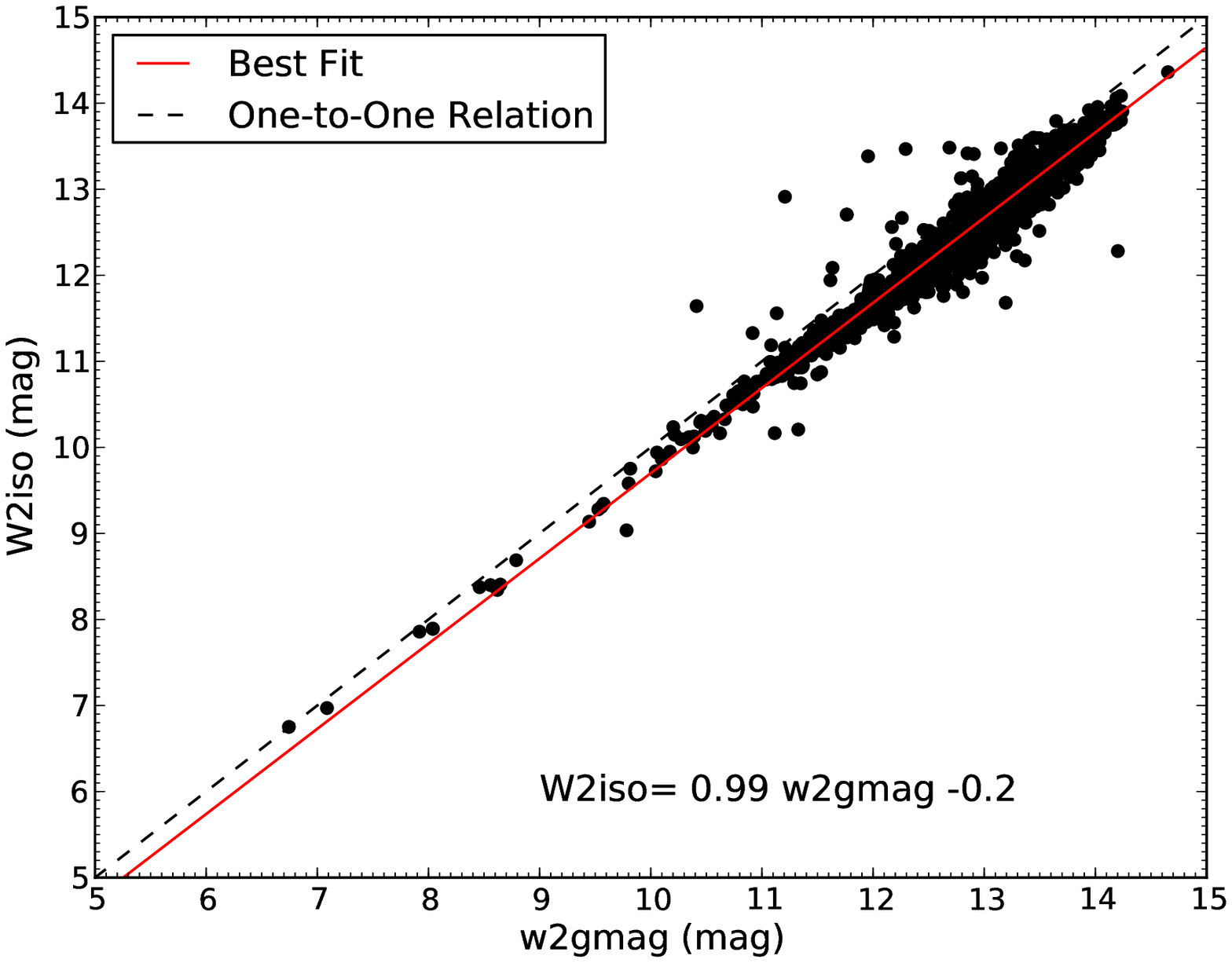}}

\caption{Comparisons of 1$\sigma$ isophotal photometry (W$\star$iso) and the 2MASS XSC-derived scaled aperture (w$\star$gmag) photometry for resolved sources in the G12 and G15 fields. See Figure \ref{resplots} of the appendix for versions plotted as a function of w$\star$gmag.}
\label{res}
\end{center}
\end{figure*}

Isolating the resolved sources using the Rfuzzy parameter (in W1) gives 1,390 and 1,368 sources in G12 and G15, respectively, i.e. $2-3$\% of the total {\it WISE} matched sources in these fields.  It is important to bear in mind that despite the resolution of the W1 and W2 bands, {\it WISE} is far more sensitive compared to, for example, 2MASS and will detect the extended light profiles from nearby galaxies. The typical W1 1$\sigma$ isophotal radius is more than a factor of $\simeq$2 in scale compared to the equivalent 2MASS K$_{\rm s}$-band isophotal radius.
An illustration of this is shown in Figure \ref{res} where the 2MASS XSC-derived scaled aperture photometry ({\it w$\star$gmag}) are compared to the isophotal photometry of resolved sources in the GAMA G12 and G15 fields. The 2MASS XSC-derived magnitudes underestimate the inferred flux due to the increased sensitivity of {\it WISE}. Note that no star subtraction or deblending has been attempted when measuring the {\it w$\star$gmag}s and contamination may be present; as a consequence, these measurements should only be used as a last resort (employing the offset shown in Figure \ref{res} to obtain the correct galaxy flux). The behavior of resolved sources is further explored in \S\,\ref{resphot} of the Appendix.

\subsection{GAMA-{\it WISE} Catalog Photometry}

For point sources, the primary photometry are the profile fit measurements ({\it w$\star$mpro}), and for well-resolved sources the isophotal photometry (\S\,3.1).  When the source is not well resolved, or the S/N (signal-to-noise) is low, the following steps are used to
choose the best photometry:


\begin{itemize}

\item {\it w$\star$mpro} photometry is used when the S/N is low, as measured by the isophotal aperture process for a given band: S/N of 10, 10, 15 and 15 for W1, W2, W3 and W4, respectively. 

\item If the Rfuzzy parameter is false (measured in the W1 band), classify source as unresolved and use {\it w$\star$mpro} photometry for bands W1 and W2.

\item If {\it w3rchi2} $<$ 2, classify source as unresolved and revert to the {\it w3mpro} photometry.

\item If {\it w4rchi2} $<$ 2, classify source as unresolved and revert to the {\it w4mpro} photometry.

\item W1 is the most sensitive {\it WISE} band. To obtain an accurate W1$-$W2 color for galaxies resolved in W1 and W2, the following steps are employed. Firstly, an accurate W1$-$W2 color is determined using a matched aperture derived from the smaller of the two isophotal radii (which is usually W2). Then, from the isophotal magnitude of the W1 band, the corresponding W2 magnitude is determined, thereby reflecting the sensitivity of the W1 band. This is also done for the W4 band using W3 in the same way as W1, if sources are resolved at these wavelengths, which in GAMA is rare (only 16 sources in G12 and G15). We note that the W1 and W3 bands cannot be matched in this way since the flux at these wavelengths is produced by different physical processes (evolved stars vs. PAH features).

\end{itemize}

Aperture corrections are applied to isophotal measurements as detailed in the Explanatory Supplement to the {\it WISE} All-Sky Data Release Products\footnote{\url{http://{\it WISE}2.ipac.caltech.edu/docs/release/allsky/expsup/ sec4$\_$4c.html$\#$apcor}}.

\begin{figure}[!thb]
\begin{center}
\includegraphics[width=8cm]{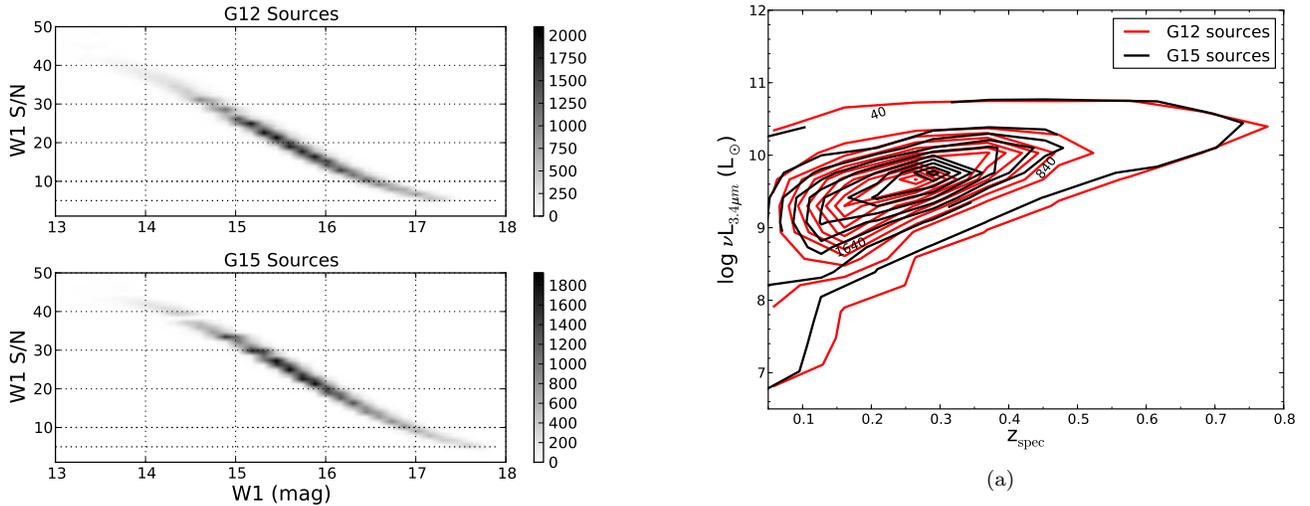}
\caption{The {\it WISE} W1 band (3.4\micron) signal-to-noise in the W1 band as a function of magnitude for G12 and G15 sources (resolved and unresolved sources). Note that G15 has better overall coverage and thus achieves greater sensitivity in luminosity.}
\label{sens}
\end{center}
\end{figure}

\begin{figure}[!ht]
\begin{center}
\subfigure[]{\includegraphics[width=8cm]{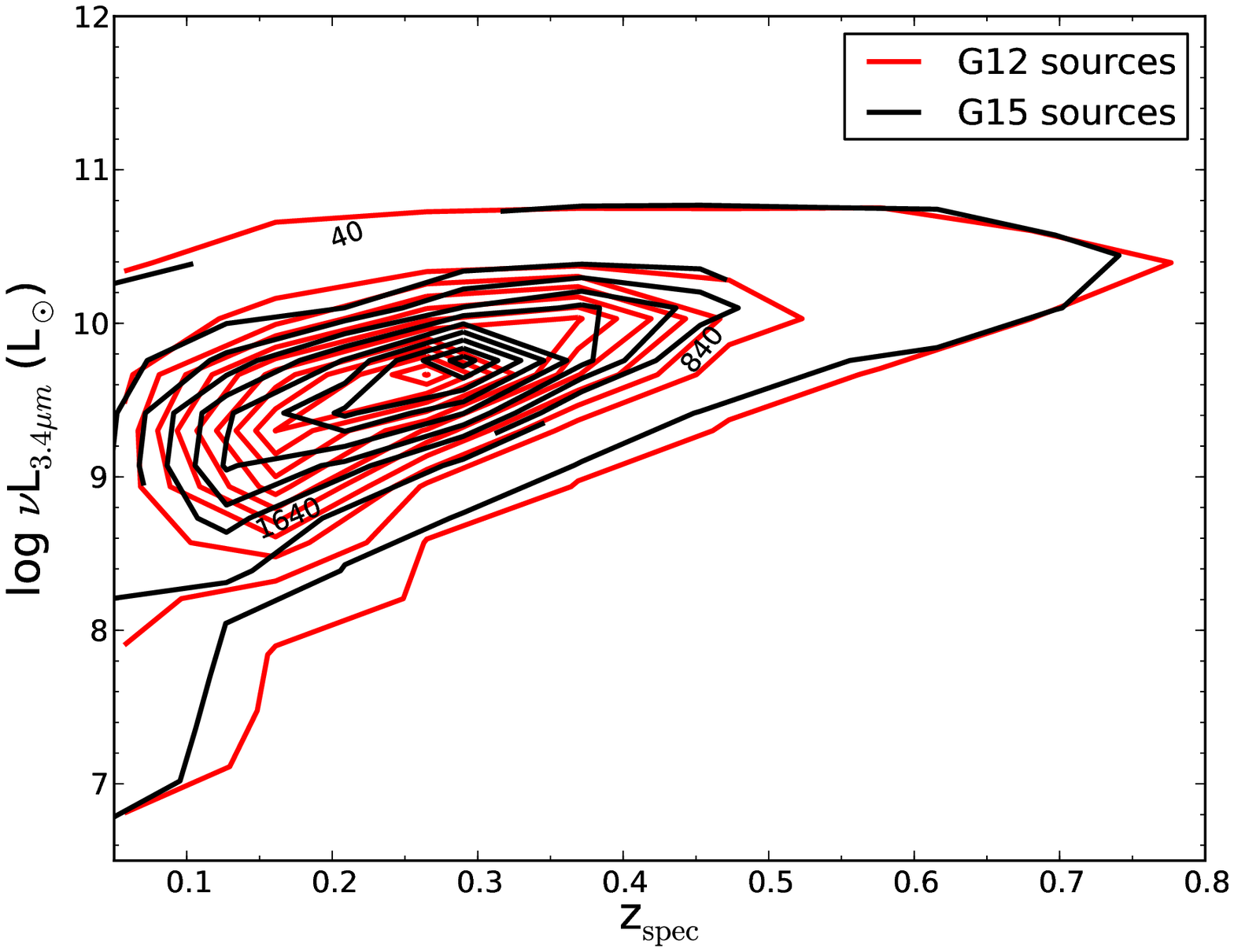}}
\hfill
\subfigure[]{\includegraphics[width=8cm]{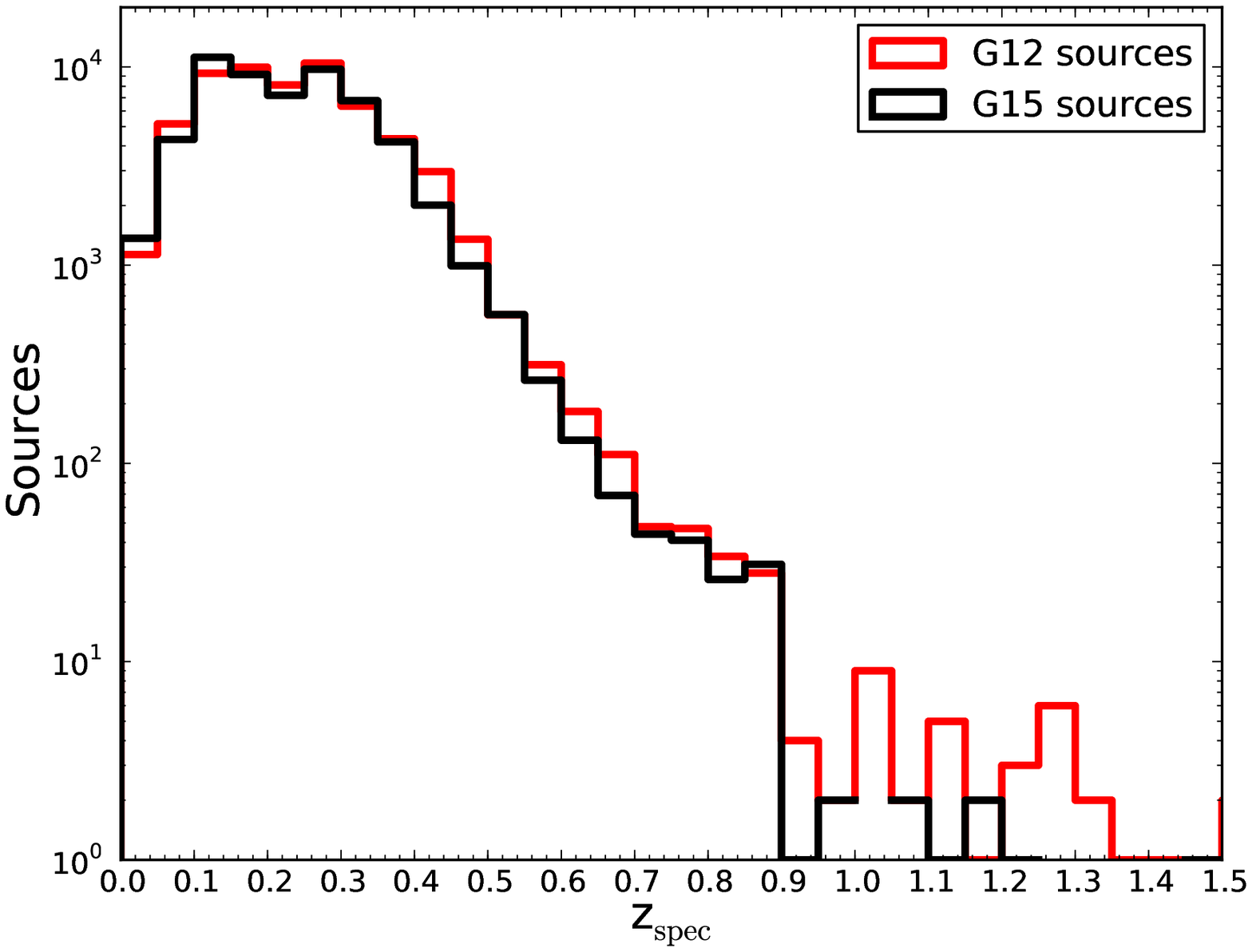}}

\caption{a) The 3.4\micron\ luminosity density ($\nu$L$_{\nu}$) distribution of sources in the two GAMA fields as a function of redshift. The number distribution as a function of redshift is shown in (b).}
\label{cat1}
\end{center}
\end{figure}

In Figure \ref{sens} we plot the signal-to-noise ratio (SNR) for the G12 and G15 fields; due to the additional $\simeq$2$\times$ frame coverage the G15 field has higher signal-to-noise detections compared to G12 at a given W1 magnitude.
The magnitude sensitivity limit in the W1 band for the G12 field is 16.6 (10-$\sigma$), 17.3 (5-$\sigma$) and for G15, 17.0 (10-$\sigma$) and 17.7 (5-$\sigma$). The combined fields have a 10-sigma magnitude sensitivity limit of 16.6 or 71 $\mu$Jy. The redshift distribution of sources in the two fields (excluding upper limits) is shown in Figure \ref{cat1}. Since the G15 field is more sensitive, the peak of the distribution is shifted to a slightly higher redshift compared to G12. The two fields show very similar redshift distributions for $z<1$ (Figure \ref{cat1}b), although the G12 matches have relatively more high-$z$ sources.

\subsection{Blending}\label{blend}

Point sources are both passively and actively deblended by the {\it WISE} pipeline and the {\it WISE} All-Sky catalog provides the {\it na} and {\it nb} blending flags where a value of ``0" indicates no active deblending has been performed, and ``1" if sources have been deblended; deblending introduces an additional uncertainty on the 
measurement provided (nb is the number of components that were deblended).

However, an additional flag is needed to indicate where {\it WISE} cannot distinguish between GAMA sources (one {\it WISE} source for multiple GAMA sources) and where contamination is expected to be high since more than one GAMA source lies within the {\it WISE} beam.  As such, we use an additional blending flag as determined by 
the proximity of neighboring GAMA-{\it WISE} sources. Sources within 5\arcs\ are viewed as a catastrophic blend due 
to the size of the {\it WISE} beam and multiple GAMA sources will probably have the 
same {\it WISE} source matched to it and/or have a highly uncertain match. A source within 15\arcs\ indicates a potential blend or contamination and is also flagged. For the analysis in this study we remove all flagged sources from our sample ($\approx$ 6\% of G12 and G15) to ensure the cleanest {\it WISE} photometry, but may bias the sample against the densest environments.

\subsection{Rest Frame Colours} 

Rest-frame colours are determined using the GAMA redshift by building an SED (Spectral Energy Distribution) combining the optical, near-infrared and mid-infrared flux densities and fitting to an empirical template library, consisting of 126 galaxy templates, of local well-studied and morphologically diverse galaxies \citep[e.g. SINGS, the {\it Spitzer} Infrared Nearby Galaxy Survey][]{Ken03} from \citet{Br13}. The templates are constructed from optical and {\it Spitzer} spectroscopy, with matched aperture photometry from {\it GALEX} \citep{Mar05}, {\it Swift} \citep[UVOT;][]{Rom05}, SDSS, 2MASS, {\it Spitzer} \citep{Wer04} and {\it WISE}, combined with MAGPHYS \citep{dacun08}. The {\it WISE} relative spectral response curves are from \citet{Jar11} and are available as part of the {\it WISE} Explanatory Supplement. 

The best template fit is determined by minimising the function that describes the difference between the flux density in the measured band, and the model-template flux density. Each resulting best-fit is assigned a normalised score derived from the reduced $\chi^2$ minimisation, with the most consistent fits having values of $<2$ and relatively poor fits with scores of $>3$ . In the sections that follow we use only those sources with a score $\le2$, although the fitting accuracy is minimally important for nearby galaxies (i.e. with low rest-frame color corrections).

\section{Results}

\subsection{{\it WISE} Colours of the G12 and G15 Fields}

\begin{figure*}[!htb]
\begin{center}
\subfigure[Observed colors]{\includegraphics[width=8cm]{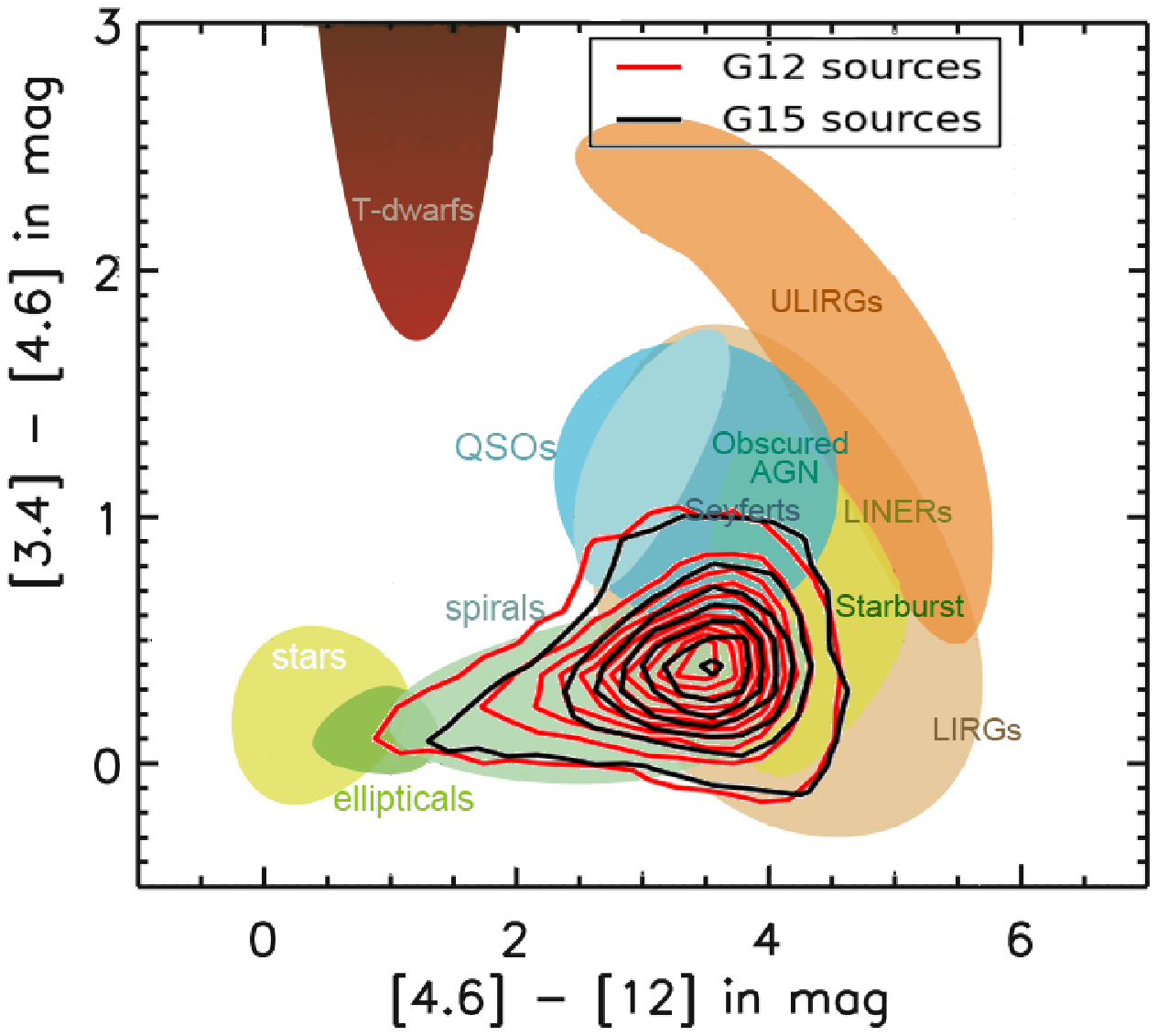}}
\hfill
\subfigure[Rest-frame colors]{\includegraphics[width=8cm]{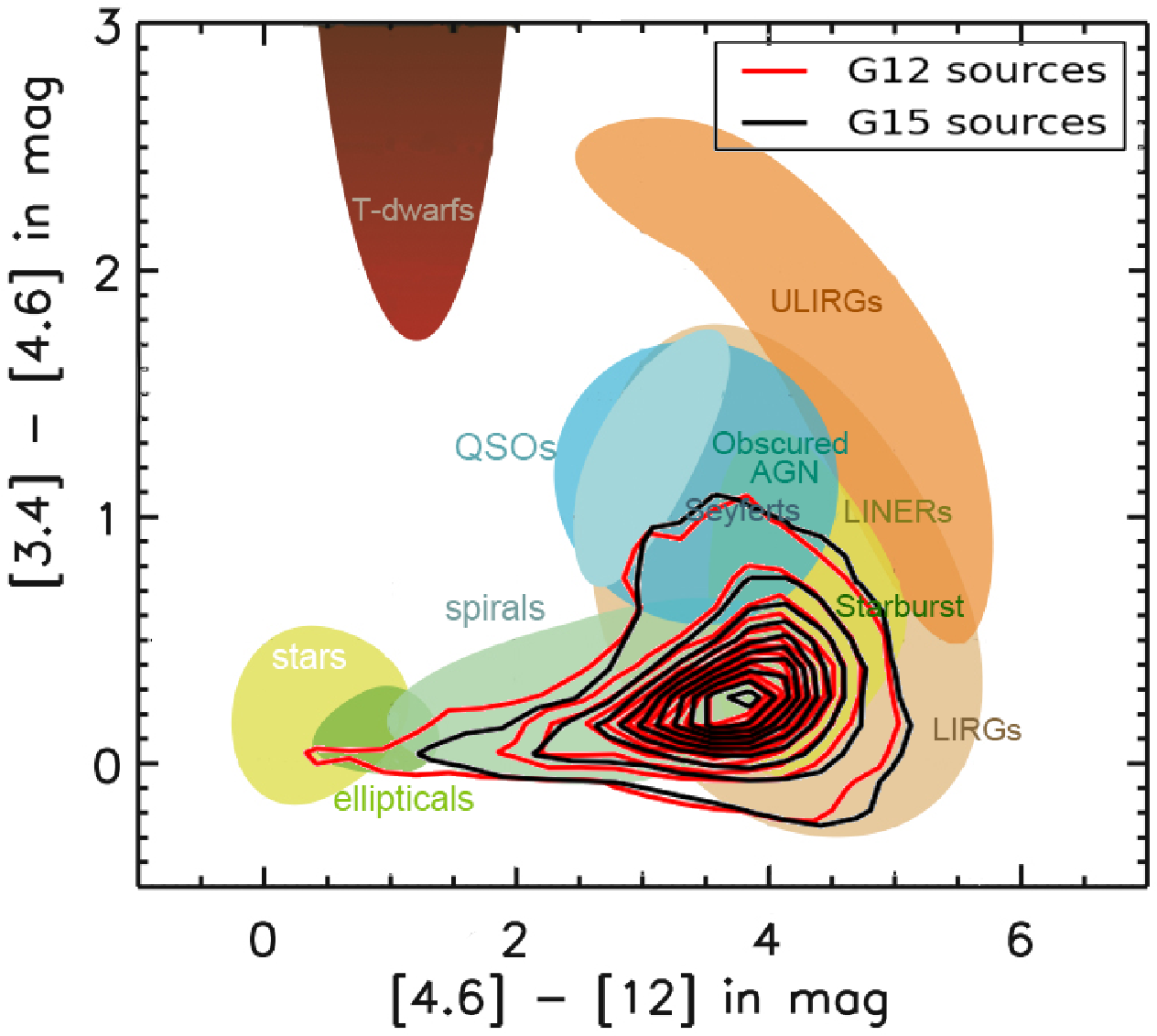}}

\caption{{\it WISE} colours of sources in G12 and G15 plotted on the color-color diagram of \citet{Jar11} with observed colors shown in (a) and rest-frame (or ``k-corrected") colors in (b).}
\label{col}
\end{center}
\end{figure*}

As shown by \citet{Jar11}, the 3.4\micron, 4.5\micron\ and 12\micron\ bands of {\it WISE} can be combined in a mid-infrared color-color diagram (Figure \ref{col}); the shortest bands are sensitive to the evolved stellar population and hot dust, therefore their color indicates increased activity (AGN or starburst). The 12\micron\ band is dominated by the 11.3\micron\ PAH, as well as the dust continuum, sensitive to star formation. This color-color diagnostic is therefore useful to separate 
galaxy populations, particularly old stellar population-dominated, star-forming and systems dominated by AGN-activity \citep{Jar11, Ster12}. In Figure \ref{col}a the observed colors of the G12 and G15 are shown, and the k-corrected version, Figure \ref{col}b, shows the distribution of the rest-frame colors. Both fields have similar distributions, biased towards star-forming systems. This is not surprising given the optical selection of the {\it GAMA} sample. Heavily obscured galaxies, notably Ultra-luminous Infrared Galaxies (ULIRGS) are absent due to the insensitivity (selective extinction) of SDSS and GAMA optical catalogs. Systems globally dominated by AGN-activity also appear sparse within the sample; this is consistent with the findings of \citet{Gun13} where spectroscopically-identified AGN make up $<20\%$ of the GAMA I emission-line catalog \citep[see also][]{Lar13}.

\subsection{Aggregate Stellar Mass Estimation}

\begin{figure*}[!th]
\begin{center}
\subfigure[Resolved Sources]{\includegraphics[height=7cm]{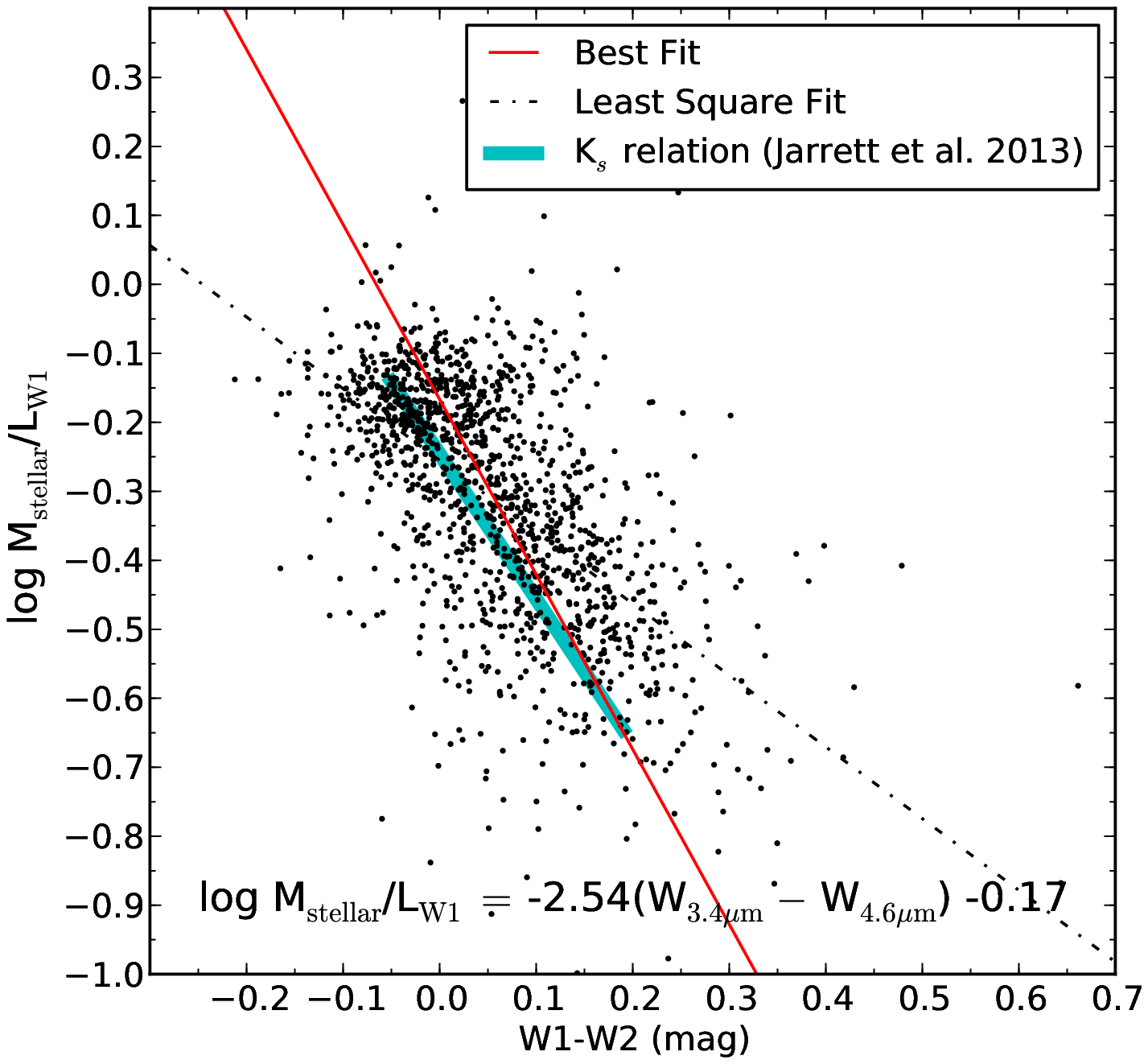}}
\hfill
\subfigure[All Sources $z<0.12$]{\includegraphics[height=7cm]{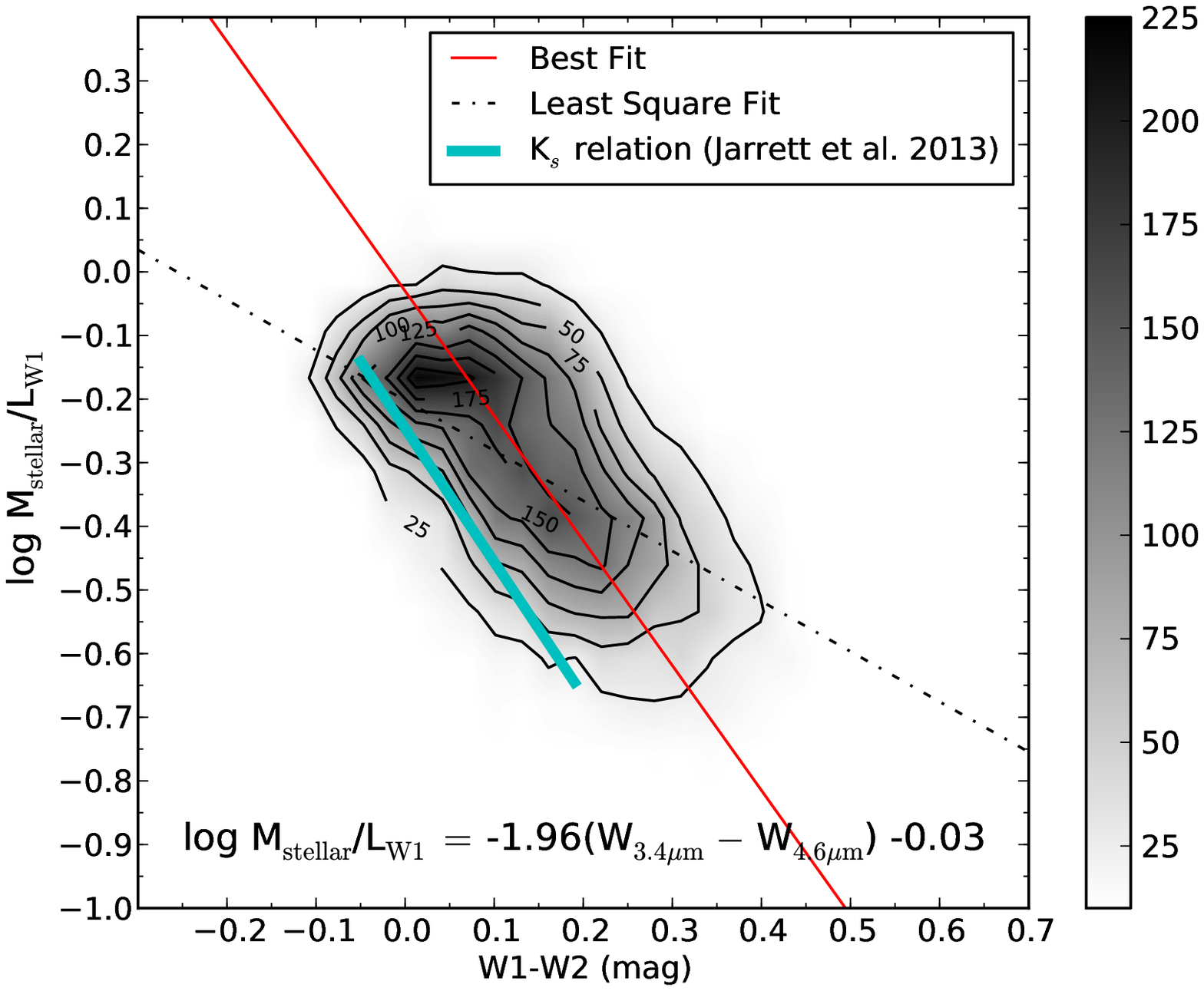}}

\caption{The 3.4\micron\ mass-to-light ratio plotted as a function of W1$-$W2 (Vega) color for a) resolved sources and b) all sources in G12 and G15. The larger volume sample shows a shift to more active galaxies with ``warmer" W1-W2 colors.}
\label{sm1}
\end{center}
\end{figure*}

The 3.4\micron\ band of {\it WISE} is dominated by the light from old stars and can be used as an effective measure of stellar mass \citep{Jar13}. To explore this further, we calculate the `in-band' luminosity for W1, i.e. the luminosity of the source as measured relative to the Sun in the W1 band, and use the stellar masses for GAMA as determined by \citet{Tay11} to determine a mass-to-light (M/L) ratio. These stellar masses are best constrained for $z< 0.15$ \citep{Tay11} and we apply this cut to our sample. As shown by \citet{Jar13}, the stellar mass-to-light has a linear trend with {\it WISE} W1$-$W2 and W2$-$W3 color, reflecting systematic M/L differences between passive and star-forming systems.

In order to empirically calibrate a relation we require high signal-to-noise measurements and additionally use only the rest-frame W1$-$W2 color since the detection rate is much higher in W2 compared to W3. We apply a S/N cut of 13.5 in W1 and W2 and remove known AGN \citep[from the GAMA I spectroscopy measurements of][]{Gun13} and also systems with {\it WISE} colors consistent with AGN activity dominating their global colors (W1$-$W2$\ge 0.8$) as discussed by \citet{Ster12}.

\begin{figure*}[!thb]
\begin{center}
\subfigure[Sources with star-forming colors (W2$-$W3$\ge 1.5$)]{\includegraphics[width=8cm]{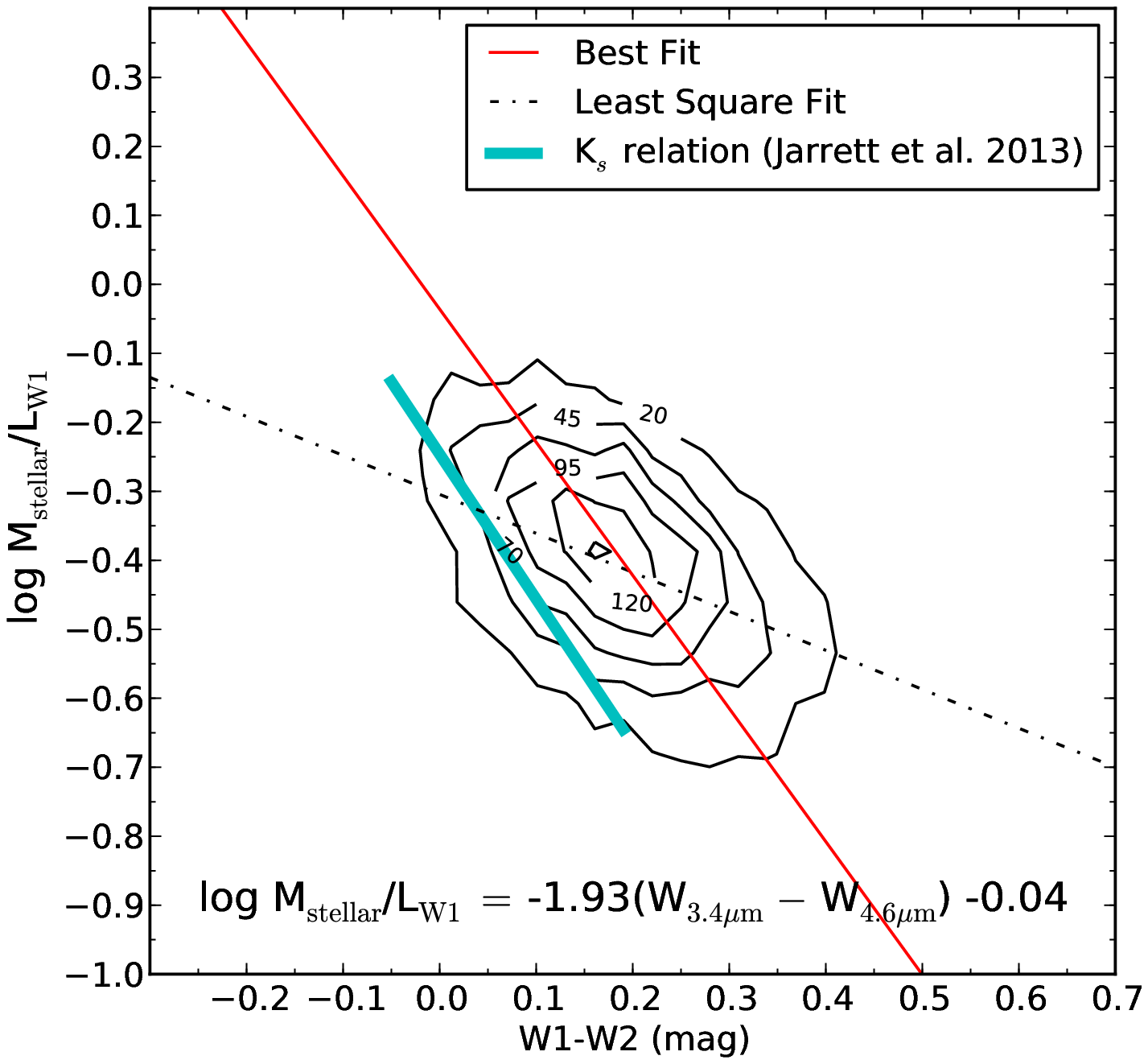}}
\hfill
\subfigure[Populations separated using the W2$-$W3 color]{\includegraphics[width=8cm]{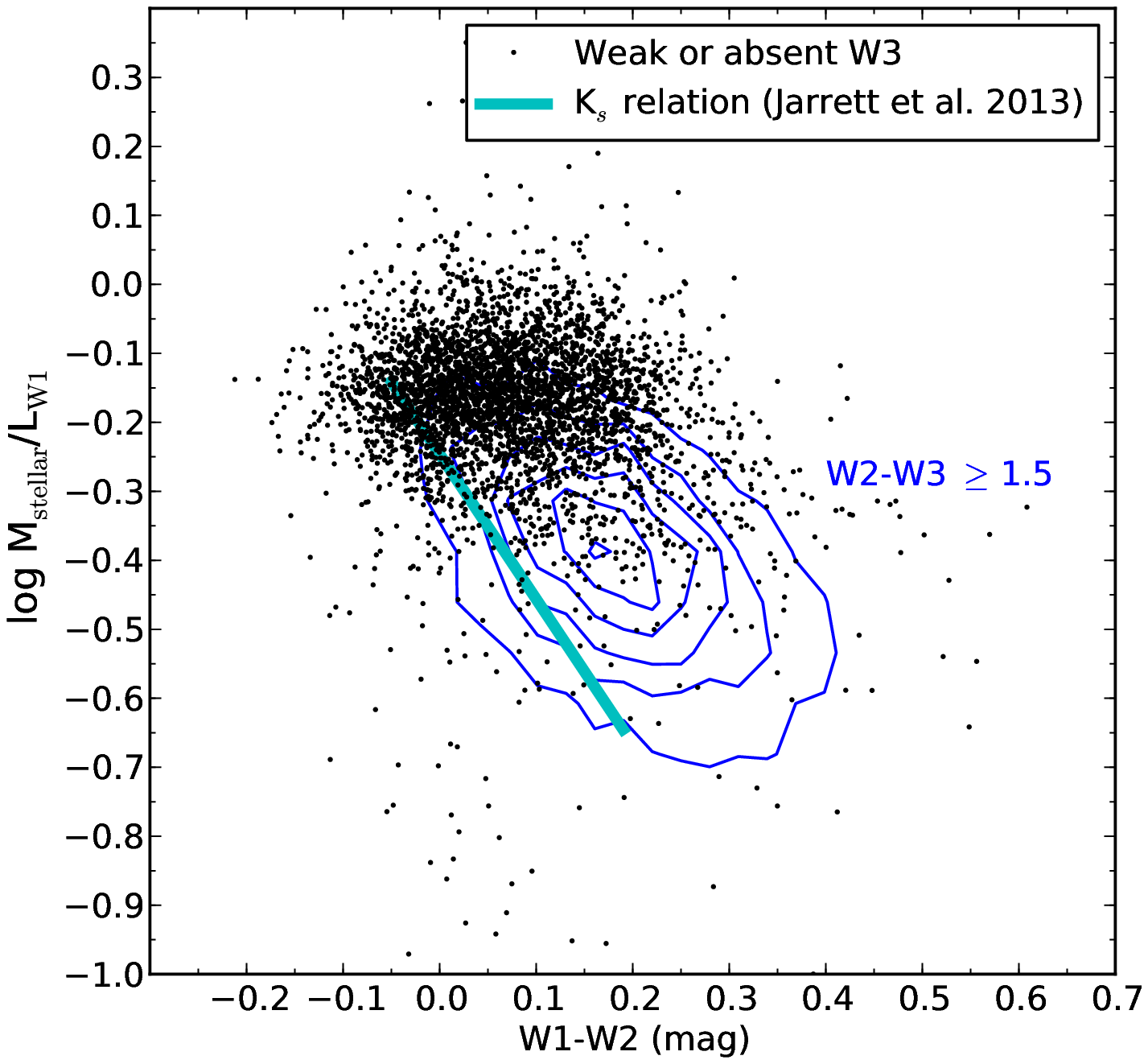}}
\caption{The W2$-$W3 color allows us to separate star-forming systems from ones that are predominantly passive.}
\label{sm2}
\end{center}
\end{figure*}

The first population we investigate are resolved, low-$z$ galaxies, 
shown in Figure \ref{sm1}a. We illustrate the danger of a least-squares minimisation fit compared to a bivariate-Gaussian, maximum-likelihood (or `best' fit) line \citep{Tay13,Hogg10}. Throughout the paper we plot both the least-squares and maximum-likelihood lines to show the large discrepancies that may exist; we provide relations for the maximum-likelihood (or best fit) only.
We find very good agreement with the M$_{\rm stellar}$/L$_{\rm W1}$ relation of \citet{Jar13} based on their relatively small sample of 17 galaxies and with stellar masses effectively derived from 2$\micron$ $K_s$ band photometry. Our relation for resolved low-$z$ sources is:

\begin{equation}
 {\rm log_{10}}\, M_{\rm stellar}/L_{\rm W1} = -2.54({\rm W}_{3.4\mu m} - {\rm W}_{4.6\mu m}) - 0.17, \\ 
 \end{equation}
 with $$  L_{\rm W1}\ ( L_{\odot}) = 10^{ -0.4 (M - M_{\rm Sun})} $$\\
where M is the absolute magnitude of the source in W1, M$_{\rm Sun} = 3.24$, and ${\rm W}_{3.4\mu m} - {\rm W}_{4.6\mu m}$ reflects the rest-frame color of the source.

For comparison to the resolved sample, in Figure \ref{sm1}b we show all sources that meet our S/N selection criterion and redshift cut, which shows the distribution to be shifted to ``warmer" W1$-$W2 colors i.e. signifying systems with increased activity such as star-formation or low-power AGN. In addition, the contours suggest that the distribution is made up of two populations, most probably due to passive galaxies having a larger mass-to-light ratio than disk-dominated and dwarf galaxies. We investigate this further in Figure \ref{sm2} where we make a color-distinction based on the W3 measurement. By selecting galaxies with W2$-$W3$\ge 1.5$ (see Figure \ref{col}) we choose systems that are most likely dominated by star formation. These selected types show a clear trend with mass-to-light (Figure \ref{sm2}a), but lie parallel and offset to the relation of \citet{Jar13}. But now including the passive galaxies, Figure \ref{sm2}b, we see a clear clustering around a fixed mass-to-light ratio of $\simeq 0.7$.

We conclude that the best-fit for our entire sample is:

\begin{equation}
 {\rm log_{10}}\, M_{\rm stellar}/L_{\rm W1} = -1.96({\rm W}_{3.4\mu m} - {\rm W}_{4.6\mu m}) - 0.03 \\ 
\end{equation}
\\
and for star-forming (lower mass-to-light) systems only:

\begin{equation}
 {\rm log_{10}}\, M_{\rm stellar}/L_{\rm W1} = -1.93({\rm W}_{3.4\mu m} - {\rm W}_{4.6\mu m}) - 0.04 \\ 
\end{equation}

This likely explains the shift observed in the relation for resolved sources and the entire field sample. The resolved sources are relatively nearby and will be a mix of passive and star-forming systems. However, at higher redshifts we add relatively more star-forming galaxies (higher infrared luminosity) causing a larger spread in W1$-$W2 color. Notably galaxies with higher star formation rates will have a larger W1$-$W2 color due to more hot dust emission giving a brighter W2 measurement. Also, at higher redshifts we will preferentially detect galaxies with higher star formation rates (see next section).  It should be noted that AGN activity would have a similar effect (as shown by the upward trend in Figure \ref{col}b) and some contamination from nuclear activity is inevitable.

\begin{figure*}[!t]
\begin{center}
{\includegraphics[width=8cm]{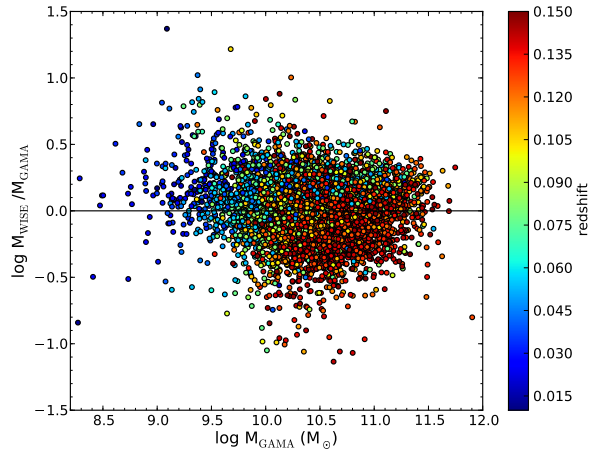}}
\caption{The difference between the WISE-derived (using equation (2)) and GAMA stellar mass estimates as a function of GAMA stellar masses \citep{Tay11}, color-coded by redshift. The mass estimates agree within a factor of 1.2 at $10^{10}$M$_\odot$ (with a standard deviation of 0.5), within a factor of 0.98 at $5\times10^{10}$M$_\odot$ (with a standard deviation of 0.4) and a factor of 0.96 at $10^{11}$M$_\odot$ (with a standard deviation of 0.4). }
\label{sm3}
\end{center}
\end{figure*}

In Figure \ref{sm3} we plot the residuals of the GAMA stellar masses (used to calibrate the WISE relation in equation 2) and the WISE-derived values themselves. The mass estimates agree within a factor of 1.2 at $10^{10}$M$_\odot$ (with a standard deviation of 0.5), within a factor of 0.98 at $5\times10^{10}$M$_\odot$ (with a standard deviation of 0.4) and a factor of 0.96 at $10^{11}$M$_\odot$ (with a standard deviation of 0.4). For stellar masses $>10^{10}$M$_\odot$ the WISE-derived masses appear overall lower than the GAMA stellar masses. This is probably unsurprising given that the sample is dominated by star-forming galaxies (see Figure \ref{col}) and the WISE 3.4\micron\ band is sensitive to the light from evolved stars in passively evolving galaxies.  

\subsection{Star Formation Rate Comparisons}

\subsubsection{22\micron\ Warm Dust Continuum}

{\it IRAS}, {\it ISO} and {\it Spitzer} revolutionised our understanding of dust emission as a tracer of star formation. In the mid-infrared, the {\it Spitzer} MIPS 24\micron\ band measures the warm dust continuum excited by hot, young stars and is therefore sensitive to recent star formation, as well as AGN-activity. Numerous studies have investigated its stability as a measure of SFR \citep[see e.g.][]{Al06, Cal07, Riek09} and we are able to transfer this understanding to {\it WISE} and its 22\micron\ band. 

\begin{figure}[!tb]
\begin{center}
\includegraphics[width=8cm]{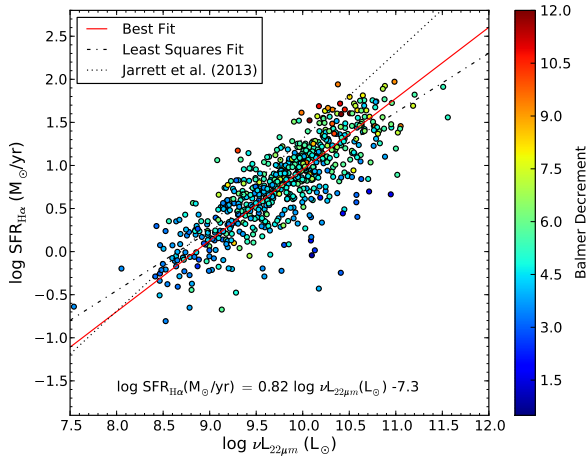}
\caption{H$\alpha$-derived star formation rates as a function of $\nu$L$_{22\mu m}$ luminosity color-coded by Balmer Decrement (H$\alpha$/H$\beta$) from \citet{Gun13}.}
\label{sf2}
\end{center}
\end{figure}

We determine the luminosity density ($\nu$L$_{\nu}$) for the 22\micron\ band and normalise by the total solar luminosity (L$_{\odot}$= 3.839$\times$10$^{26}$\,W). The W4 band is the least sensitive {\it WISE} band, but we impose a S/N cut of 7 to ensure high quality photometry. Sources flagged as AGN based on optical spectroscopy diagnostics are removed from the sample. Cross-matching these sources with star formation rates available for GAMA I, using sources with SFR between 0.1 and 100 M$_{\odot}$\,yr$^{-1}$ \citep[see ][]{Gun13}, yields the best-fit relation shown in Figure \ref{sf2}. The optical star formation rates are determined from H$\alpha$ equivalent widths applying an extinction-correction based on the Balmer decrement \citep[full details can be found in][]{Gun13} and are sampled to $z<0.35$ (i.e. beyond which H$\alpha$ is redshifted out of the observed spectral range). Galaxies where the H$\alpha$ line is contaminated by atmospheric O$_2$ (A band) absorption, in the redshift range 0.155$< z < $ 0.170 are not included in the sample as recommended in \citet{Gun13}. The points in Figure \ref{sf2} are color-coded according to Balmer decrement and on average higher infrared luminosities correspond to higher Balmer decrements (i.e., increased dust obscuration). However, without detailed knowledge of the dust geometry this is merely a crude comparison \citep[see also][]{Wij11a, Wij11b}, and as we discuss below, biases due to extinction are probably in play at higher redshifts. Balmer decrement has also been shown to be correlated with stellar mass, stellar mass surface density and metallicity \citep[see for example,][]{Bos13}

Figure \ref{sf2} also shows the linear fit of \citet{Jar13}, where the {\it WISE} 22\micron\ -based SFR relation was calibrated using  {\it Spitzer} 24\micron\ photometry and the relation of \citet{Riek09}. This fit is somewhat steeper, likely explained by the distinction made by \citet{Riek09} based on total infrared luminosity of the source and the relatively small number of sources in the \citet{Jar13} sample (but which also included Local Group dwarf galaxies). The best fit to the GAMA sample distribution is:

\begin{equation}
 {\rm log_{10}\, SFR}_{{\rm H}\alpha} (M_\odot/{\rm yr}) = 0.82\, {\rm log_{10}}\, \nu L_{22\mu \rm m} (L_\odot) -7.3 \\
\end{equation}

\subsubsection{12\micron\ ISM Tracer}

Infrared-luminous sources will be detected by {\it WISE} 22\micron\ out to moderate redshifts ($z\sim$ 2 to 3), but most typical galaxies would not. Hence there is a strong bias for W4 detections to be luminous galaxies (LIRGS and ULIRGS), which are highly obscured at optical wavelengths.
The {\it WISE} 12\micron\ band, on the other hand, has greater sensitivity by comparison (see Table \ref{match}) and also probes the ISM, sensitive to a larger (representative) sampling of galaxies, and thus 
making it potentially the primary star formation indicator for {\it WISE}. 
The dominant feature within this W3 band is the 11.3\micron\ PAH (polycyclic aromatic hydrocarbon), and to a lesser extent the $[$Ne{\sc ii}$]$ emission line. This PAH is large, neutral and excited by ultraviolet radiation from young stars, as well as radiation from older, evolved stars \citep[see, for example,][]{Kan08}. $[$Ne{\sc ii}$]$ is associated with emission from \HII\ regions. Passive disks generally lack the 7.7\micron\ PAH tracing current star formation, but still show prominent 11.3\micron\ PAH features \citep[see, for example,][]{Clu13}.  As first demonstrated with {\it Spitzer} measurements \citep[see for example][]{Hou07,Farr07}, PAHs can be used to estimate star formation, although there is larger scatter (and potential biases) relative to the superior 24\micron\ mid-infrared or 70\micron\ far-infrared tracers. The {\it WISE} 12\micron\ band is thus a powerful, yet also more problematic, tracer of recent star formation.

\begin{figure}[!thb]
\begin{center}
\subfigure[]{\includegraphics[width=8cm]{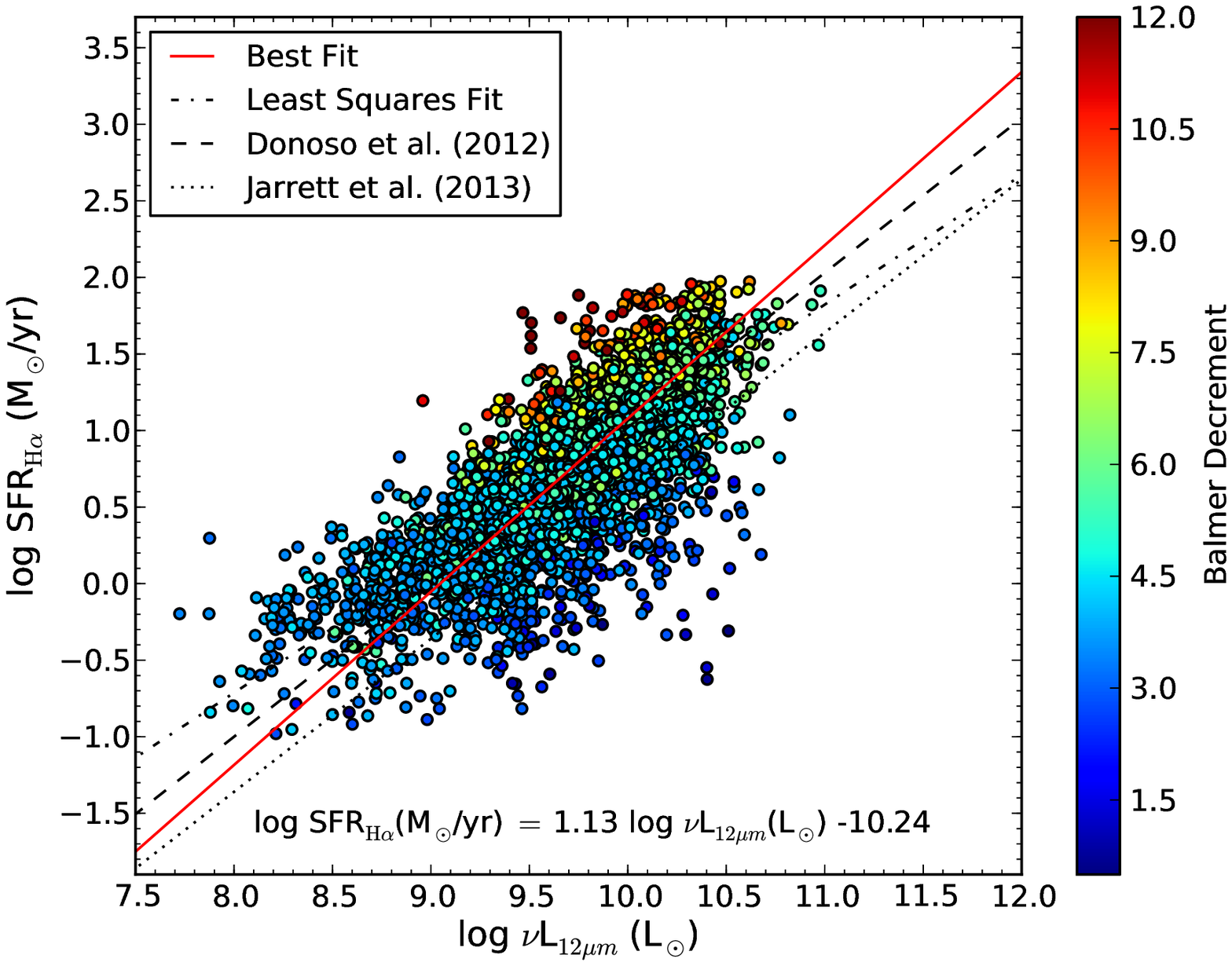}}
\hfill
\subfigure[]{\includegraphics[width=8cm]{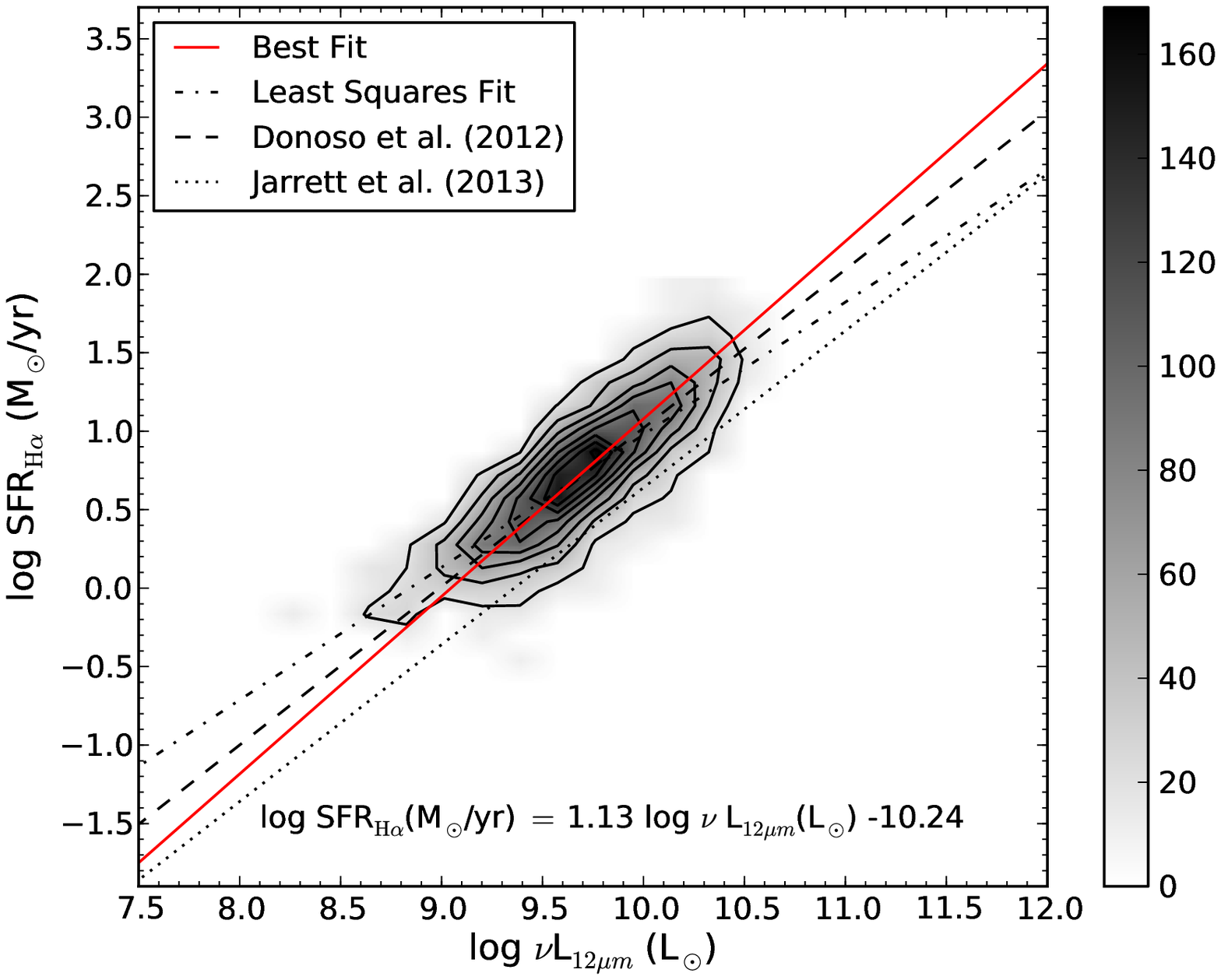}}

\caption{H$\alpha$-derived star formation rates as a function of $\nu$L$_{12\mu m}$ luminosity color-coded by Balmer Decrement in (a) and in (b) shown as a contour and density plot.}
\label{sf1}
\end{center}
\end{figure}

We proceed as with W4, but with a S/N cut of 10 in the W3 (12\micron\ band), similarly color-coding to reflect Balmer decrement (Figure \ref{sf1}a). A contour plot of the distribution is shown in Figure \ref{sf1}b and the trend appears very tight for SFR $>5\, M_\odot\,{\rm yr}^{-1}$. At low SFR and $\nu$L$_{\nu}$, however, the distribution appears to flatten and probably accounts for the differences in slope between our relation and the least squares fit of \citet{Don12}.
The relation of \citet{Jar13} lies offset below our best fit, most likely as a result of relatively low SFR within their small sample (SFR $<5\, M_\odot\,{\rm yr}^{-1}$) of nearby galaxies. The best fit relation for the 12\micron\ band is:

\begin{equation}
 {\rm log_{10}\, SFR}_{{\rm H}\alpha} (M_\odot/{\rm yr}) = 1.13\, {\rm log_{10}}\, \nu L_{12\mu \rm m} (L_\odot) -10.24 \\
\end{equation}

This flattening of the distribution in Figure \ref{sf1} at low SFR is most likely a feature of the 11.3\micron\ PAH tracer (since we do not observe it at 22\micron) and is also observed by \citet{Lee13} in their study.

Given that the 12 micron band emission appears too low for the given H$\alpha$ SFR, a possible explanation is that the relative abundance of PAH molecules to big grain emission is diminished, due to low metalicity in these galaxies. If this is the case, we do not expect to see a similar effect at 22 micron, since this band is dominated by big grains in equilibrium with the strong radiation fields inside star formation regions. Indeed, the 22\micron\ sample of \citet{Lee13} does not show such a flattening. 

An alternative explanation would be that in a low SFR regime, the preponderance of the diffuse medium increases with respect to the star-formation component of the interstellar medium (galaxies are more quiescent). Bear in mind that the 24\micron\ warm dust emission is powered by the UV radiation fields from the massive stars inside the star forming regions, while the 12\micron\ emission is mainly powered by the diffuse interstellar radiation fields \citep{Pop11}. Thus, the PAH and small grains emission seen in the 12\micron\ band increases with increasing SFR, but this increase is mediated by the propagation of the UV photons in the diffuse ISM, while the 24\micron\ emission directly traces star formation regions, as does the Balmer line corrected H$\alpha$
emission from which the SFRs are derived.



\begin{figure*}[!htb]
\begin{center}
\subfigure[]{\includegraphics[width=8cm]{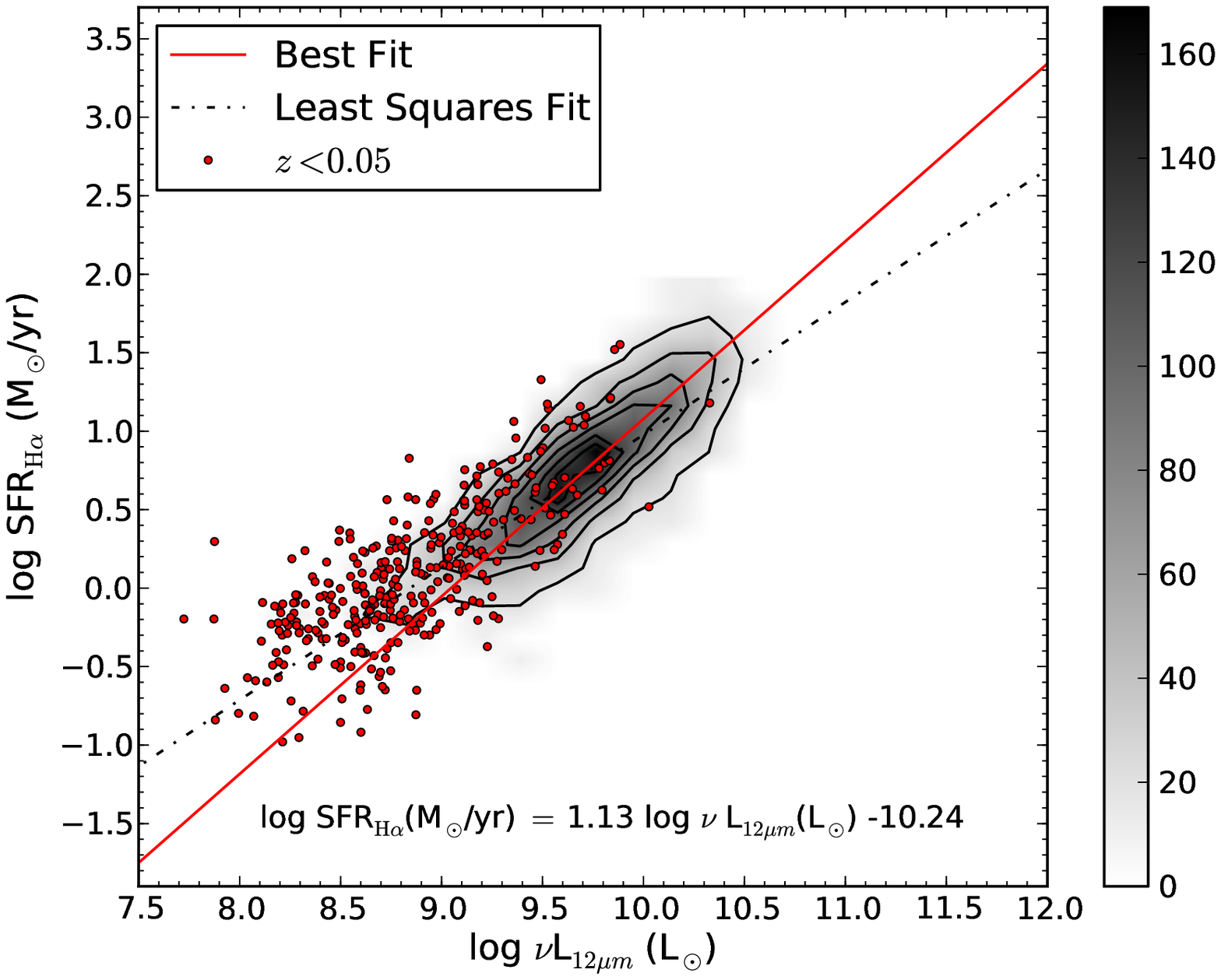}}
\hfill
\subfigure[]{\includegraphics[width=8cm]{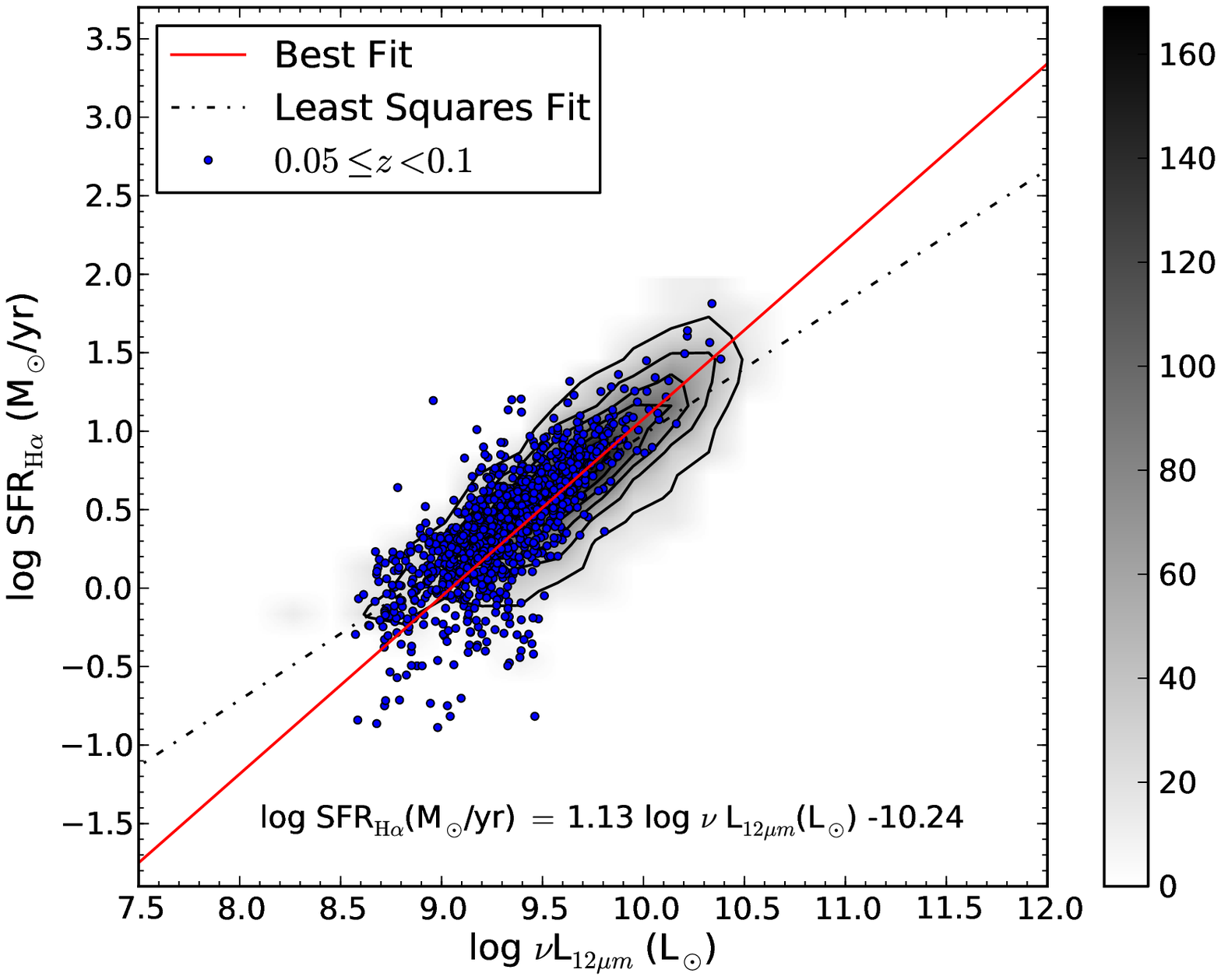}}
\vfill
\subfigure[]{\includegraphics[width=8cm]{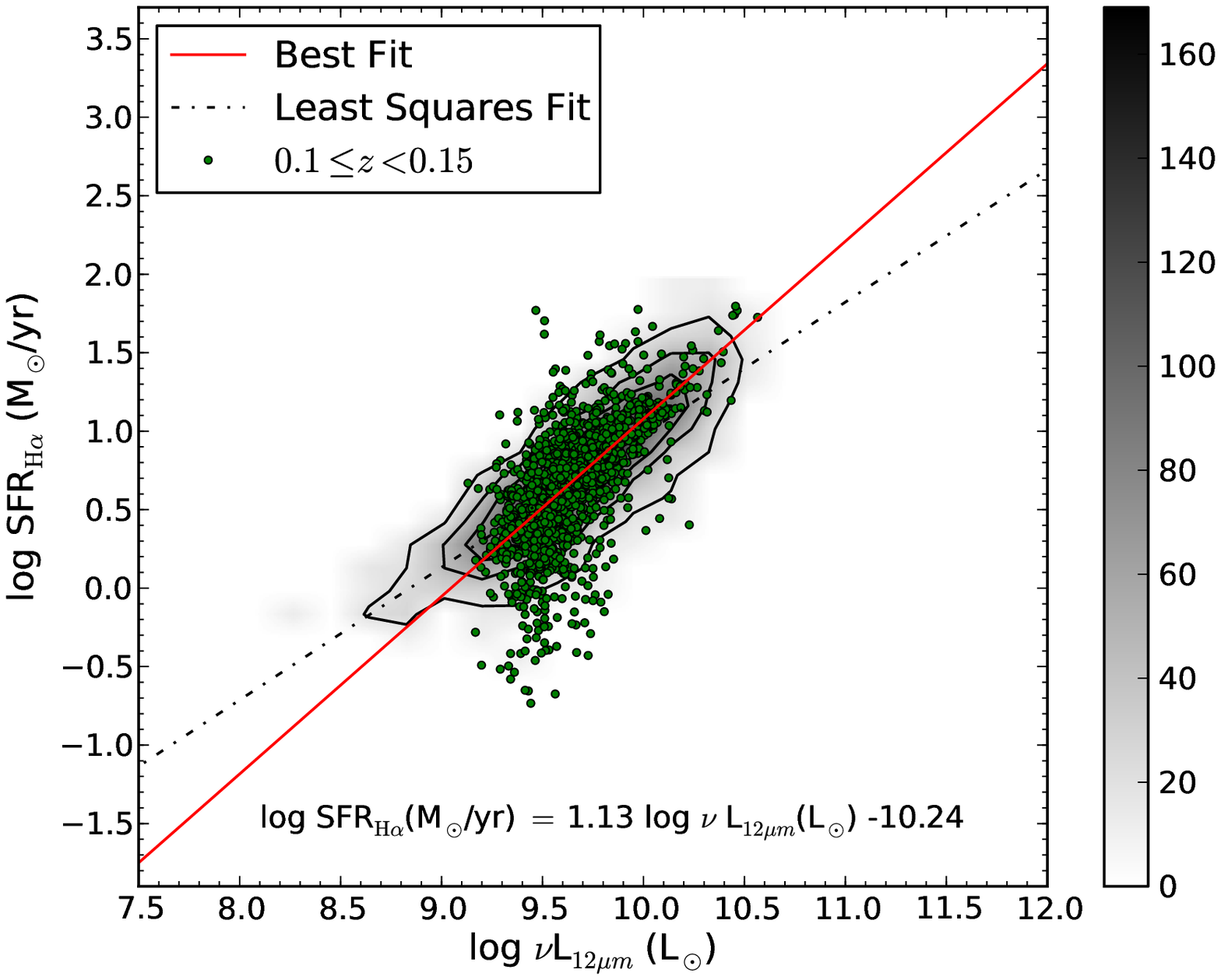}}
\hfill
\subfigure[]{\includegraphics[width=8cm]{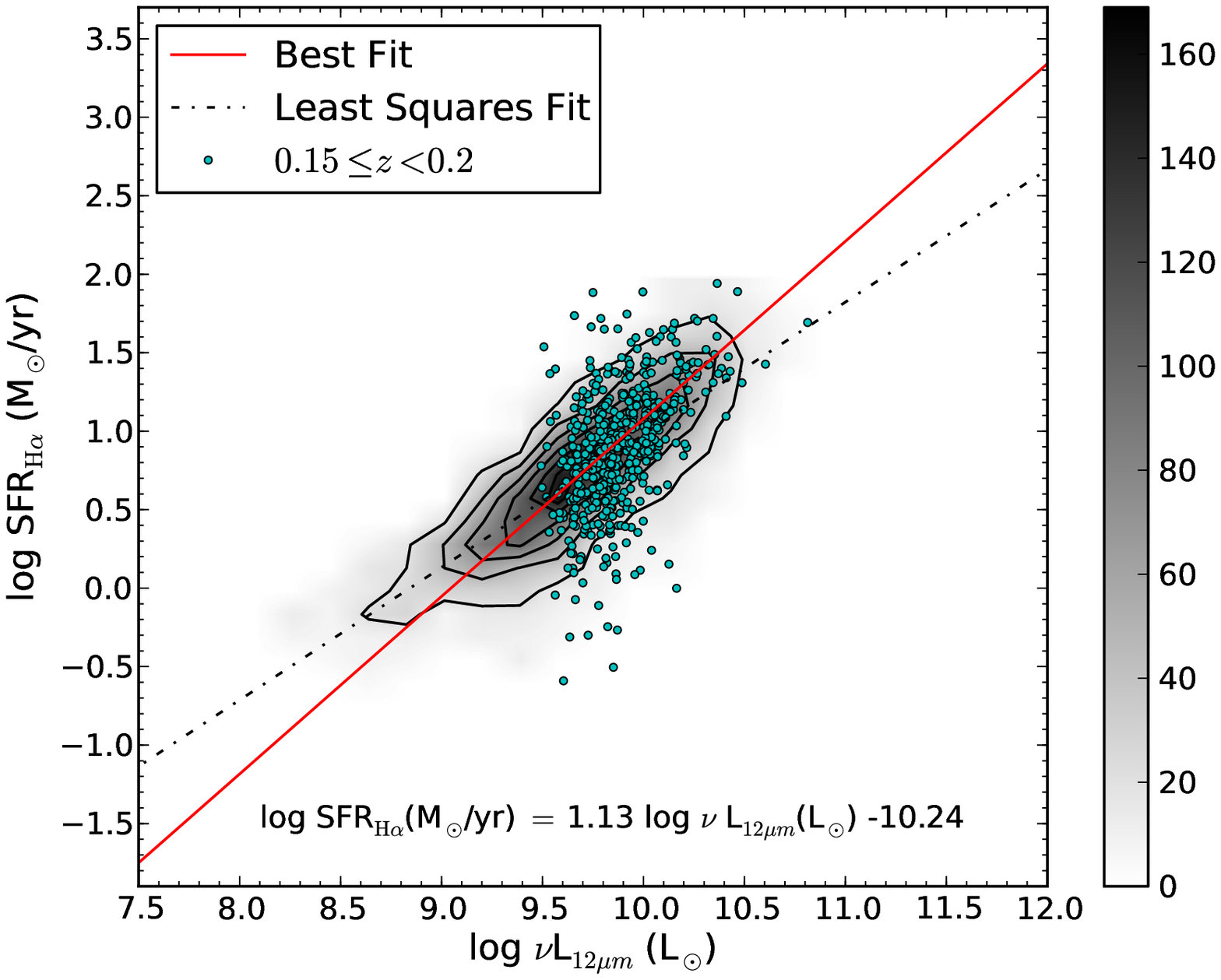}}

\caption{H$\alpha$-derived star formation rates as a function of $\nu$L$_{12\mu m}$ luminosity for redshift slices of 0.05 between $0<z<0.2$ overplotted on the full distribution from Figure \ref{sf2}.}
\label{sf3}
\end{center}
\end{figure*}

To investigate this further we show individual points for contiguous redshift slices overplotted on the full sample in Figure \ref{sf3}; this serves as an indication of how the distribution is built up. The flattening at low SFR appears to be a feature of nearby sources alone (Figure \ref{sf3}a) supporting the hypothesis that we are observing the PAH features and small grain emission of relatively quiescent galaxies, dominated by the diffuse ISM. The flattening in the slope is not detected at higher redshifts ($z>0.05$) as these galaxies have more star formation activity (i.e. more clumpy). In Figures \ref{sf3}b-d we see how as we move to higher redshifts, we sample systems with higher $\nu L_{12\mu \rm m}$ and, on average, higher SFRs. The relatively low number of sources in Figure \ref{sf3}d reflects the atmospheric contamination affecting sources at redshifts of $z\simeq0.16$.

The higher infrared luminosity in these sources, however, also indicate greater obscuration in the optical bands;  this explains the sharp drop in H$\alpha$ luminosities relative to W3, apparent in Figs. 10c and d with SFR(H$\alpha$) $<$ 3 M$_{\odot}$\,yr$^{-1}$.  Corrections using the Balmer decrement become ineffective when the extinctions are high, (Av $>>1$), which is why infrared tracers of obscured star formation and UV/optical tracers of unobscured star formation are best combined to estimate the total star formation rate.

\section{Science with the GAMA-WISE Catalog}

The GAMA-{\it WISE} catalog and empirically-derived relations can be used any number of ways to probe the behaviour of large populations of galaxies. To illustrate, we highlight below how {\it WISE} can be used to study the specific star formation of galaxies.

\subsection{Specific Star Formation}

\begin{figure}[!htb]
\begin{center}
\includegraphics[width=8cm]{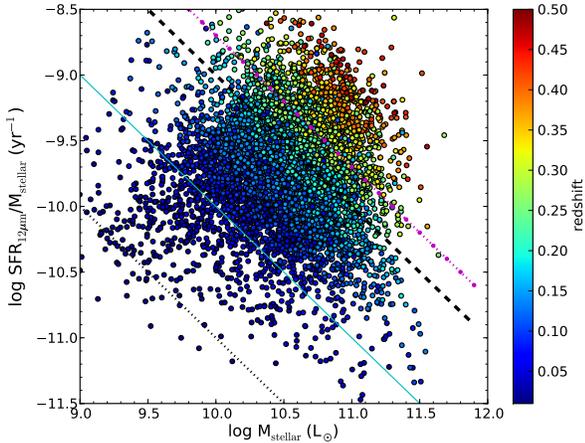}
\caption{Specific star formation with redshift.  The star formation derived from the $\nu$L$_{12\mu m}$ luminosity, for the entire sample $\nu$L$_{12\mu m}$ as a function of stellar mass, giving the sSFR,
color-coded by redshift. Lines of constant SFR (0.1, 1, 10 and 20 M$_{\odot}$\,yr$^{-1}$) are shown as dotted, solid, dashed and dash-dot lines, respectively.}
\label{ssfr}
\end{center}
\end{figure}

We can investigate the specific star formation rate (sSFR) of the entire sample by using $L_{12\mu \rm m}$-derived star formation rates (equation 5) and the GAMA stellar masses \citep{Tay11}. In Figure \ref{ssfr} we show specific star formation as a function of stellar mass, color-coded by redshift.  Two clear trends are seen: (1) lower mass galaxies are actively building their disks while massive galaxies have expended their gas reservoirs rendering mostly passive evolution (the ``SFR-M$_\star$" relation)
and (2) with increasing redshift, a shift to higher SFR for a given stellar mass as we capture these infrared-luminous systems. The behavior of star formation and specific star formation, as a function of stellar mass and redshift, within the GAMA sample is explored in detail in \citet{Bau13} and \citet{Lar13}. The observed increase in mean sSFR of star-forming galaxies with increasing redshift is well-established \citep[see for example][]{Noes07, Elb07, Rod10} and we illustrate here that the GAMA-{\it WISE} sample is sufficiently large and diverse to explore galaxy evolution between the local universe and $z<$ 0.5.

\begin{figure}[!hbt]
\begin{center}
\subfigure[]{\includegraphics[width=8cm]{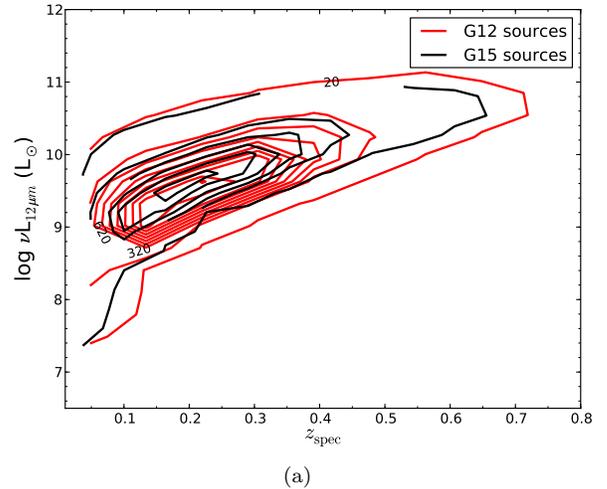}}
\hfill
\subfigure[]{\includegraphics[width=8cm]{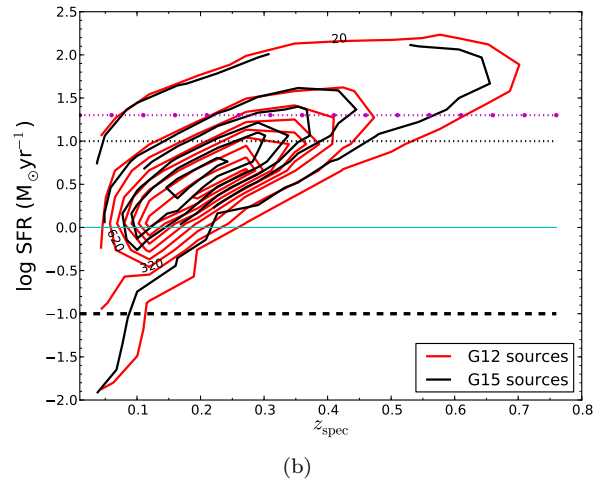}}

\caption{The redshift distribution of sources in the two GAMA fields relative to (a) the 12\micron\ luminosity density ($\nu$L$_{\nu}$) and in (b) converting luminosity density to SFR using equation (5). Lines of constant SFR (0.1, 1, 10 and 20 M$_{\odot}$\,yr$^{-1}$) are shown as dotted, solid, dashed and dash-dot lines, respectively.}
\label{lumw3}
\end{center}
\end{figure}

From the relation of \citet{Bou10}, the ``main-sequence" of galaxies with stellar mass of 10$^{11}$\,M$_\odot$ at $z\simeq0.5$ have typical SFRs of $\simeq 20$ M$_\odot$yr$^{-1}$. Converting to luminosity density by way of equation (5) gives log$_{10}L_{12\mu \rm m} \simeq 10.2$\,L$_\odot$. As illustrated in Figure \ref{lumw3}, {\it WISE} can detect these systems that are within the GAMA-{\it WISE} sample. A luminosity density of log$_{10}L_{12\mu \rm m} \simeq 10.2$\,L$_\odot$ corresponds to a flux density of 0.24 mJy or 12.8\,mag; using the S/N detection statistics of the G15 sources, {\it WISE} will detect this source with S/N$\simeq 20$. The magnitude sensitivity limit in the W3 band for the G12 field is 13.2 (10-$\sigma$), 14.0 (5-$\sigma$) and for G15, 13.8 (10-$\sigma$) and 14.5 (5-$\sigma$).

\subsection{Specific Star Formation of Galaxies in Pairs}

\begin{figure}[!hbt]
\begin{center}
\subfigure[Pair galaxies]{\includegraphics[width=8cm]{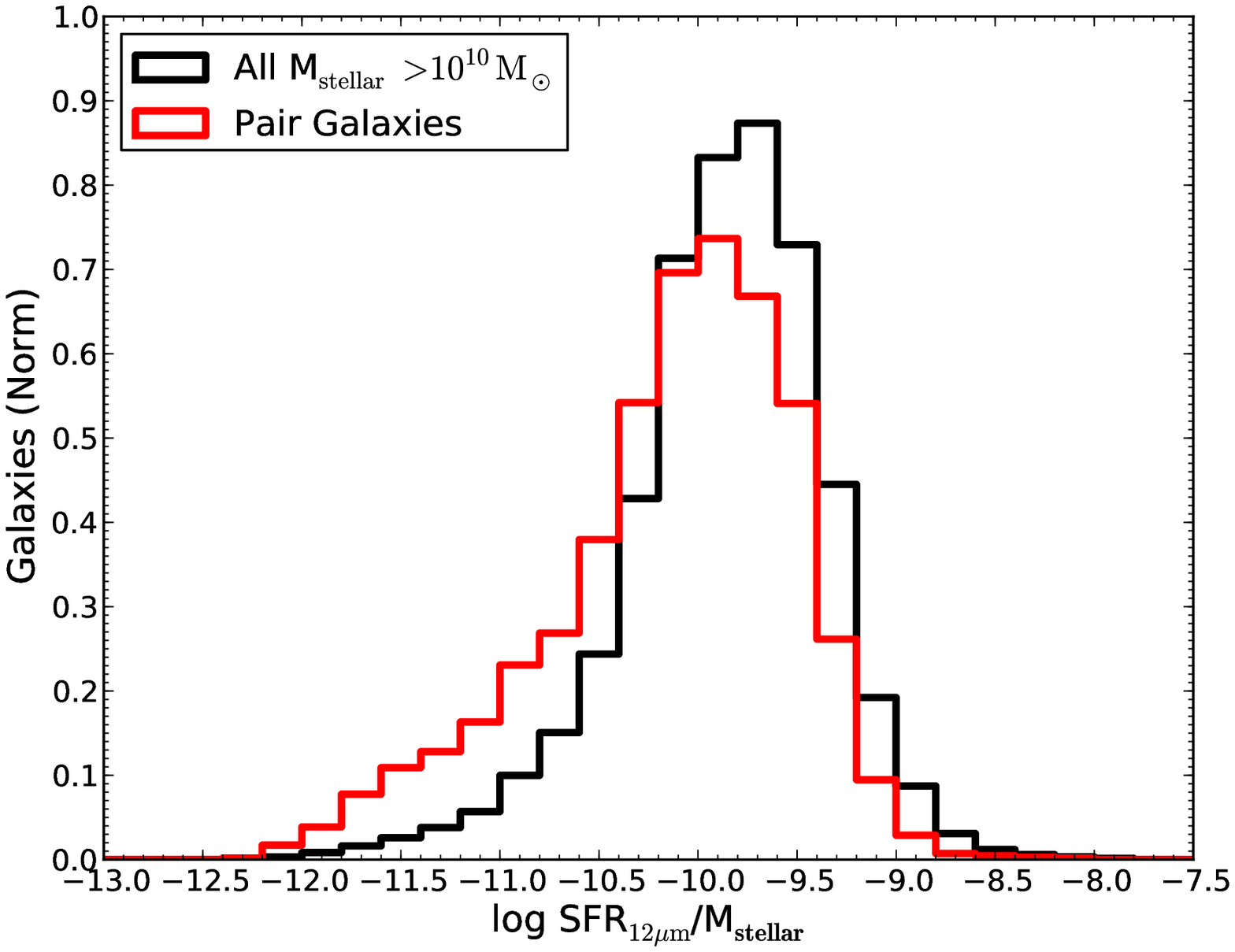}}
\hfill
\subfigure[Pair and `close pair' galaxies]{\includegraphics[width=8cm]{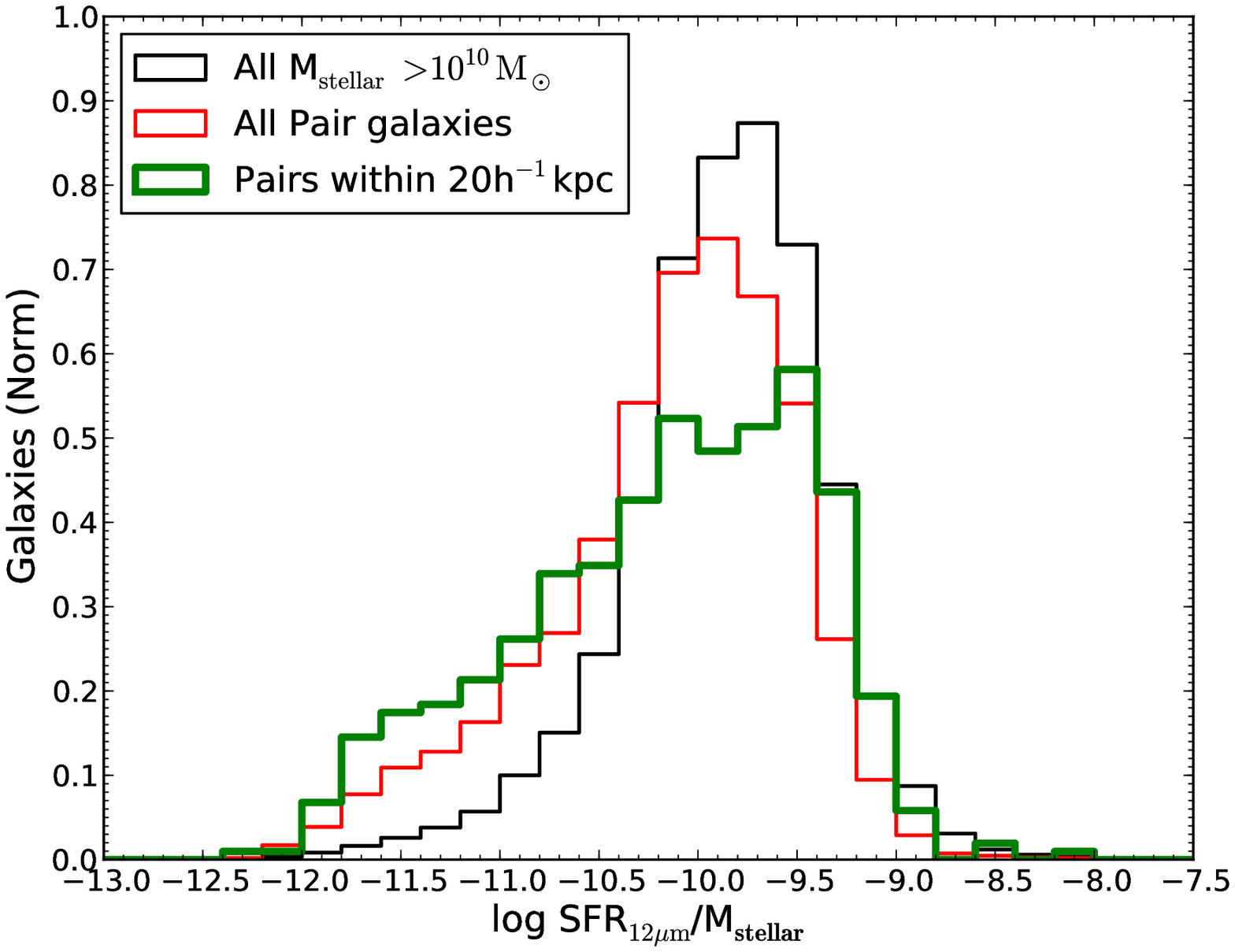}}
\caption{Specific star formation rate normalised distribution of galaxies with stellar masses $>10^{10}$M$_{\odot}$ and the subsets that reside within a pair and a close pair (separation $<$ 20\,h$^{-1}$\,kpc).}
\label{pairs}
\end{center}
\end{figure}

The 12\micron-derived star formation rates, computed from equation (5), can also be used to investigate sSFR trends within populations of GAMA galaxies. For example, we use the updated GAMA II Galaxy Group Catalogue (G$^3$Cv6), as detailed in \citet{Rob11}, to isolate galaxies that reside within a pair. In G$^3$C a galaxy pair is defined as two galaxies within 100\,$h^{-1}{\rm kpc}$, in physical separation, and 1000 km\,s$^{-1}$ in velocity separation. Figure \ref{pairs} is a normalised histogram of all galaxies in the sample with stellar mass $>$ 10$^{10}$M$_{\odot}$ and the subset that reside in a pair (17,475 galaxies) and a close pair (i.e. separation $<20\,h^{-1}{\rm kpc}$). We note that potential blends (as outlined in \S\,\ref{blend}) are removed.

The striking feature of Figure \ref{pairs}a is that galaxies that reside within a pair appear to have, on average, lower sSFR than a typical GAMA galaxy of the same mass, suggesting that instead of star formation being enhanced in these systems, it is suppressed.  Although initially counter-intuitive, and contrary to studies of local galaxy pairs compared to field control samples \citep[see, for example,][]{Ell13}, it should be noted that $\sim40\%$ of GAMA galaxies reside within a group, with numerous mechanisms at play. Within the context, therefore,  of the environment of a typical galaxy this highlights the complexity of interacting and merging systems and is discussed further in a detailed study of merging and interacting galaxies within GAMA (Robotham et al., in prep.). We note that a lack of SFR enhancement has emerged from studies of galaxy pairs probing higher redshift samples \citep[see, for example, ][]{Xu12}, with one suggested explanation that higher gas fractions at higher redshift reduce the efficiency of torque-driven gas infall.

Further we include in Figure \ref{pairs}b a histogram of galaxies within 20\,kpc/h of their neighbor, as compared to the distribution from Figure \ref{pairs}a. It appears that this subset shows a broader range of sSFRs compared to the larger sample of galaxies in pairs, and may even be bimodal, consisting of suppressed systems and enhanced systems.  Adding this kinematic trigger to galaxy evolution is clearly a complex process, but crucial to understand how galaxies evolve in the group environment. Finally, we note that blending limitations of the {\it WISE} data within pairs that have a smaller separation make this parameter space uncertain, but optical tracers will be exploited here in forthcoming GAMA papers.

\section{Conclusions}

We have detailed the procedure for creating a source-matched catalog between galaxies in the GAMA G12 and G15 fields, and {\it WISE} photometry. In particular we have outlined how best to extract photometry for low signal to noise, unresolved and resolved sources in the {\it WISE} All-Sky Catalog. Complete GAMA-{\it WISE} catalogs for the G09, G12 and G15 fields will be made available through the GAMA Public Releases (see http://www.gama-survey.org/).

Using the {\it WISE} measurements and matched GAMA galaxies in the G12 and G15 fields, we have investigated the following:

\begin{itemize}

\item The {\it WISE} color distribution for the sample shows most systems are globally dominated by star formation, with few passive and AGN-dominated systems; this is consistent with the GAMA sample selection and spectroscopic sensitivity to higher redshift, star-forming galaxies. 

\item Empirical relations of stellar mass as a function of 3.4\micron$-$4.6\micron\ color for resolved sources, our entire sample and star formation-dominated galaxies only. We provide relations that can be applied to large samples for $z<0.5$.

\item Star formation rate relations can be derived using 22\micron\ and 12\micron\ luminosities. The 12\micron-derived SFR relation (equation 5) is a complex tracer of the ISM and should, however, be used with caution. We show that the distribution of galaxies in the 12\micron\ luminosity and H$\alpha$ SFR plane as a function of redshift are affected by selection effects and most probably dust geometry. 

\item Specific star formation (using 12\micron-derived SFRs) for the sample illustrates how the GAMA-{\it WISE} catalog detects only the most massive, highest star-forming systems at the highest redshifts. {\it WISE} is, however, able to detect star-forming main-sequence systems, of stellar mass $\sim$10$^{11}$M$_\odot$, to $z\simeq 0.5$ with S/N $>10$. 

\item Controlling for stellar mass, galaxies with an associated neighbor appear to experience, on average, a shift to lower specific star formation. Extracting pairs with relatively small separations ($<$ 20\,$h^{-1}$kpc) suggests a broader behaviour consistent with populations experiencing either star formation suppression or enhancement. 

\end{itemize}

\section*{Acknowledgements}

MEC and THJ acknowledge support from the National Research Foundation (South Africa). MEC, MLL, MO, AEB and MJIB acknowledge support from the Australian Research Council (FS110200023, FS100100065, FT100100280). MLPG acknowledges support from a European Research Council Starting Grant (DEGAS-259586). 

This publication makes use of data products from the Wide-field Infrared Survey Explorer, which is a joint project of the University of California, Los Angeles, and the Jet Propulsion Laboratory/California Institute of Technology, funded by the National Aeronautics and Space Administration.

GAMA is a joint European-Australasian project based around
a spectroscopic campaign using the Anglo-Australian Telescope.
The GAMA input catalog is based on data taken from the
Sloan Digital Sky Survey and the UKIRT Infrared Deep Sky Survey.
Complementary imaging of the GAMA regions is being obtained
by a number of independent survey programs including
GALEX MIS, VST KiDS, VISTA VIKING, WISE, Herschel-
ATLAS, GMRT and ASKAP providing UV to radio coverage.
GAMA is funded by the STFC (UK), the ARC (Australia), the
AAO, and the participating institutions. The GAMA website is
http://www.gama-survey.org/.

\appendix

\section{Analysis of {\it WISE} Photometry for Nominally Resolved Sources}

In order to understand the photometry of resolved and partially resolved sources in {\it WISE} we use several diagnostics. Since {\it WISE} is calibrated in the Vega system, images and all-sky catalogs, all comparisons are done as such.

\subsection{Rfuzzy Defined}\label{rfuzz_sec}

The Rfuzzy parameter is used to determine whether a source is resolved in the 3.4\micron\ band of {\it WISE} (see \S\,\ref{rfuzzy}).
The Rfuzzy parameter is measured in the following way:

\begin{enumerate}

\item The source is rotated to determine the optimal 2nd moment parameters (Rmajor, Rminor, position angle of major axis)

\item The source is divided along the minor axis into two halves:  positive x, and negative x, where the central x position is the nominal position of the source

\item Compute the 2nd order intensity-weighted moment for each half and derive the Rmajor for each half: Rmajor$_{\rm a}$, Rmajor$_{\rm b}$

\item Rfuzzy $=$ minimum(Rmajor$_{\rm a}$, Rmajor$_{\rm b}$).

\end{enumerate}

\begin{figure}[hpt]
\begin{center}
\subfigure[Rfuzzy as a function of W1 isophotal signal to noise]{\includegraphics[width=8cm]{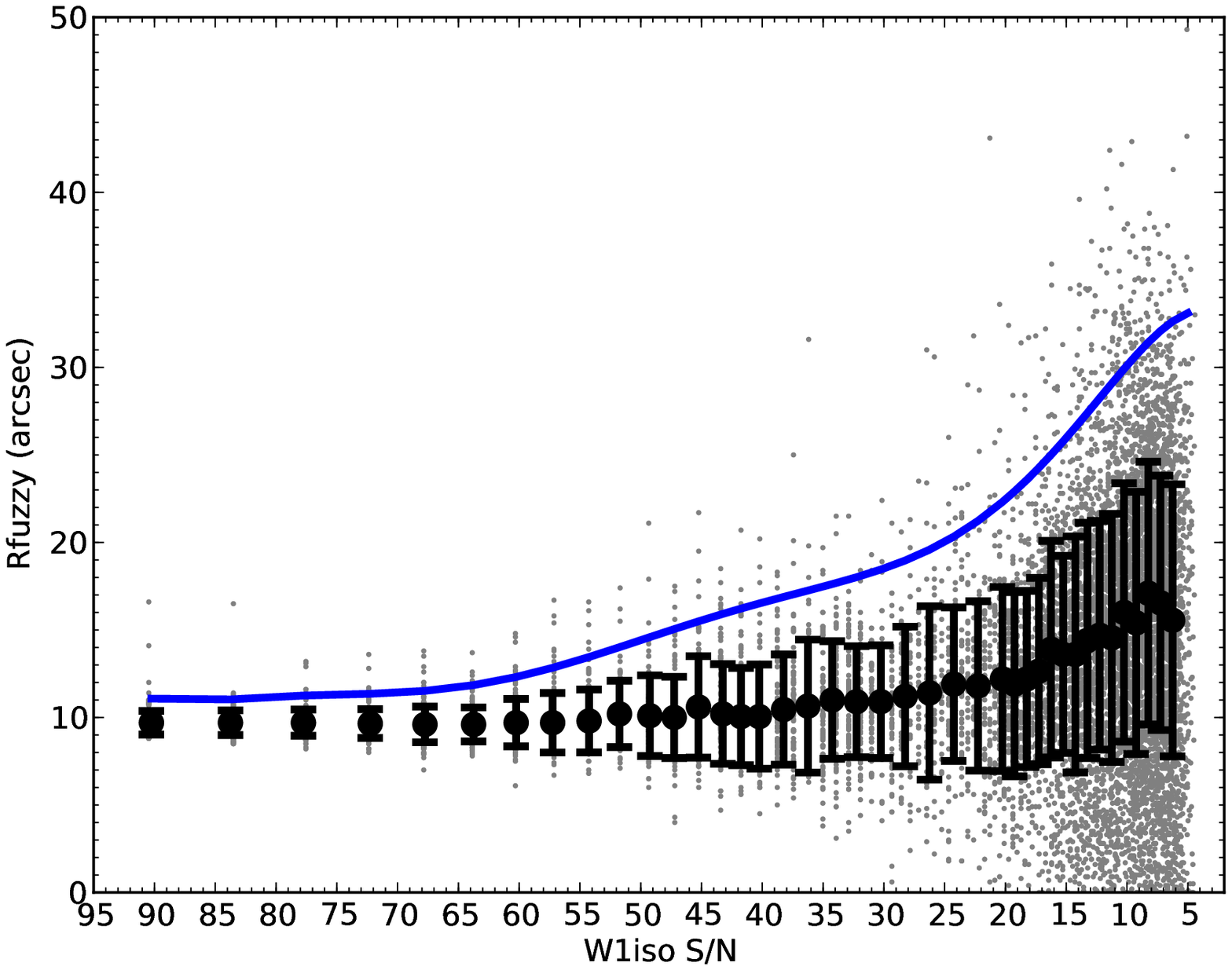}}
\subfigure[Rfuzzy as a function of W1 isophotal magnitude]{\includegraphics[width=8cm]{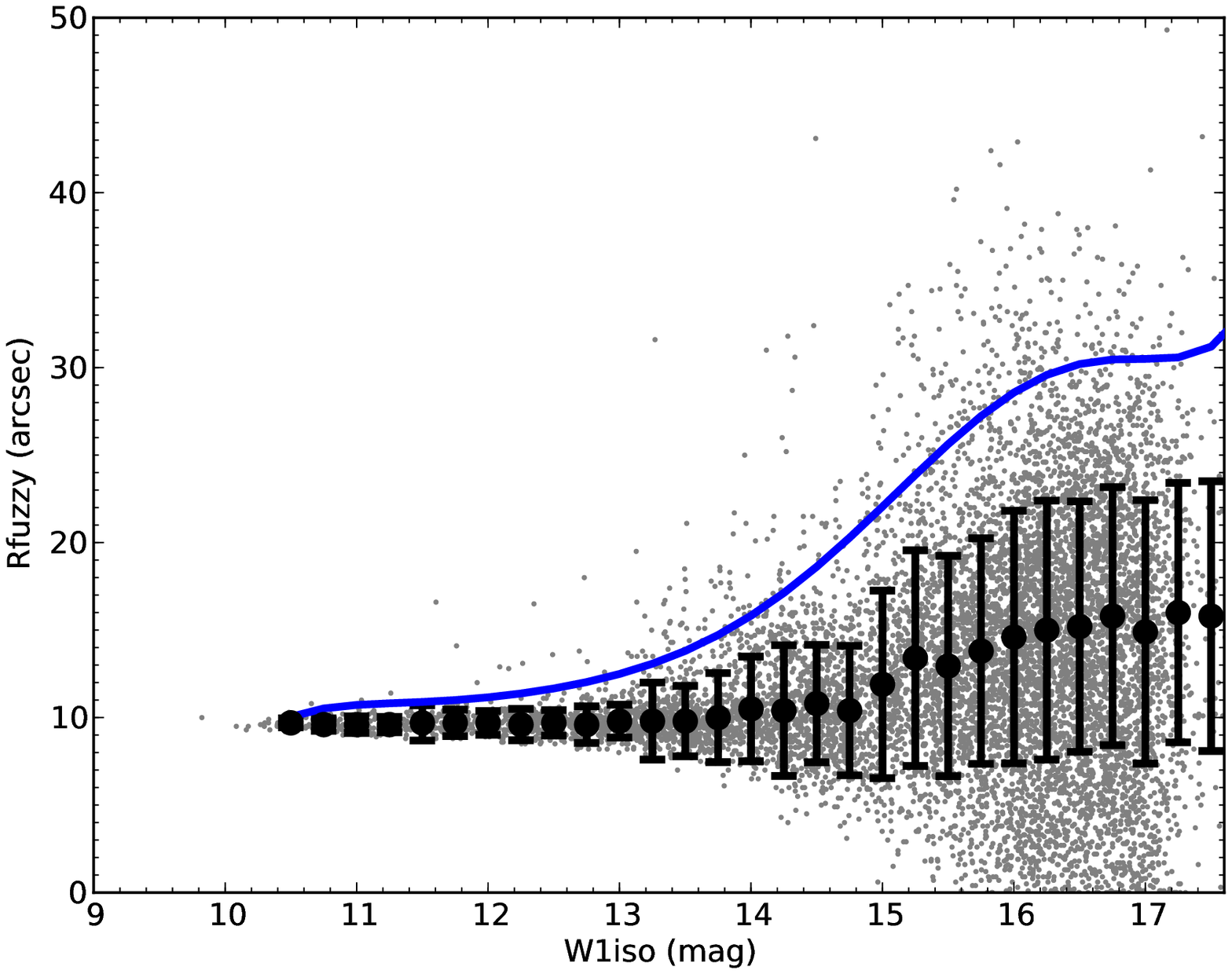}}
\caption{Rfuzzy$_{\rm limit}$ is calibrated using sources in G12 that are expected to be unresolved i.e. {\it w1rchi2}$<$2.}
\label{rfuzz}
\end{center}
\end{figure}

If Rfuzzy $<$ Rfuzzy$_{\rm limit}$, then the source is unresolved. The Rfuzzy$_{\rm limit}$ is determined by measuring sources that are expected to be unresolved. For example, Figure \ref{rfuzz} shows the Rfuzzy values for {\it WISE} sources in the GAMA G12 region with {\it w1rchi2}$<$2 which preferentially selects unresolved sources. 
The typical value for Rfuzzy is $\simeq$10.0\arcs\ for high signal to noise point sources (e.g. stars), but the distribution can be used to identify the limit where resolvedness can be determined. A power series fit to the 2$\sigma$ mean of the distribution yields the Rfuzzy$_{\rm limit}$ curve -- points that lie above this curve are classified as resolved.

Rfuzzy can be defined in terms of isophotal signal to noise or isophotal magnitude. Given the differing exposure coverage (number of {\it WISE} frames that samples a given patch of sky) for the entire sky, the sensitivity limits vary depending on location on the sky. For this reason, a magnitude-dependent Rfuzzy$_{\rm limit}$ curve will be sensitive to the {\it WISE} coverage and would need to be derived for each region. However, a signal to noise function is robust against this and therefore the curve shown in Figure \ref{rfuzz}a is used. 

\begin{figure}[hpt]
\begin{center}
\includegraphics[width=8cm]{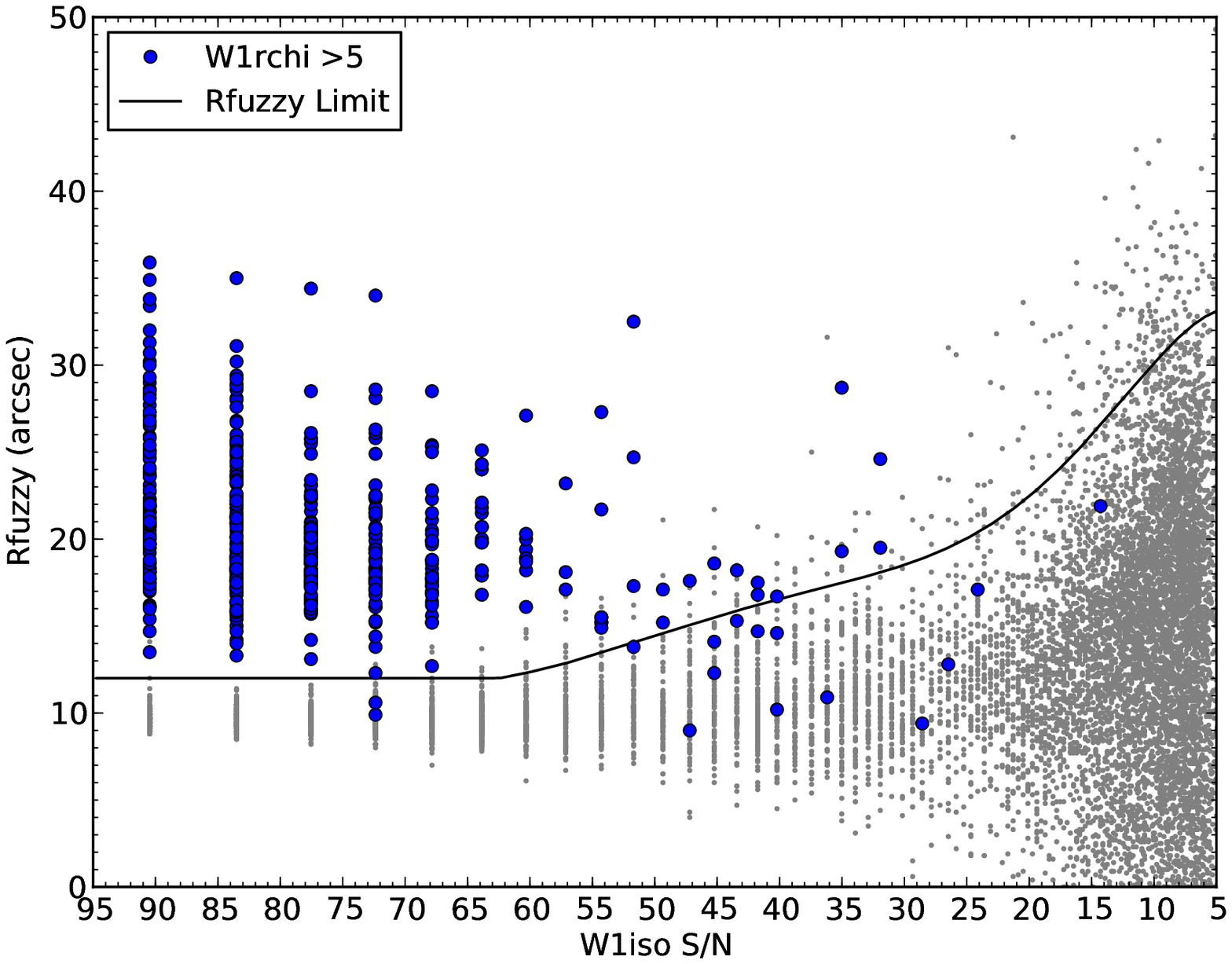}
\caption{The performance of Rfuzzy is shown for sources in G15 where the majority are expected to be resolved i.e. {\it w1$\star$rchi2}$>$5.}
\label{rfuzz_perf}
\end{center}
\end{figure}

As a test of the performance of the Rfuzzy parameter, Figure \ref{rfuzz_perf} plots Rfuzzy for a sample from GAMA G12 expected to be dominated by resolved systems (based on {\it w1rchi2}). The majority of sources with {\it w1rchi2}$>$5 lie above the curve defined by the point sources in Figure \ref{rfuzz}a.

\begin{figure*}
\begin{center}
\subfigure[]{\includegraphics[width=8cm]{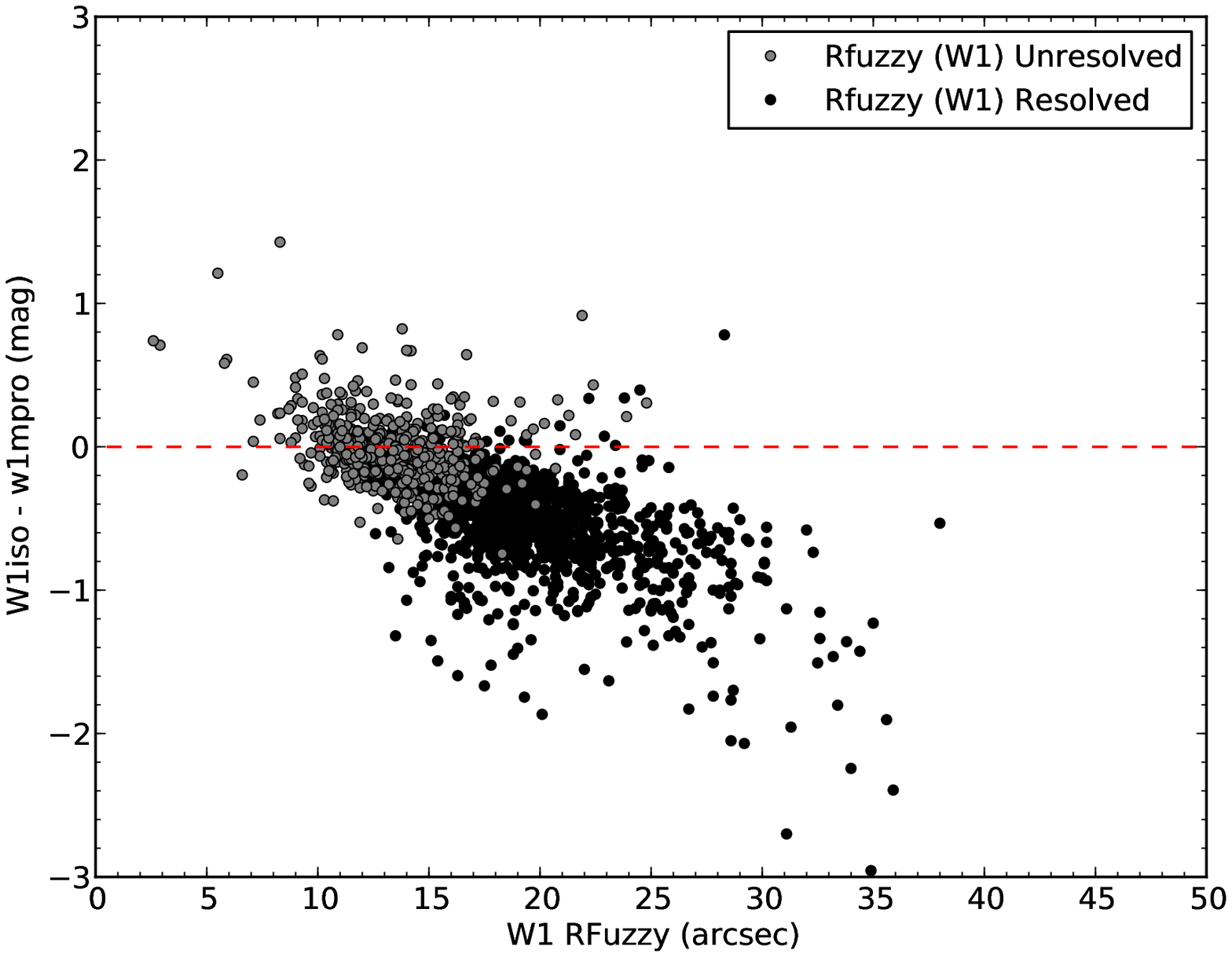}}
\hfill
\subfigure[]{\includegraphics[width=8cm]{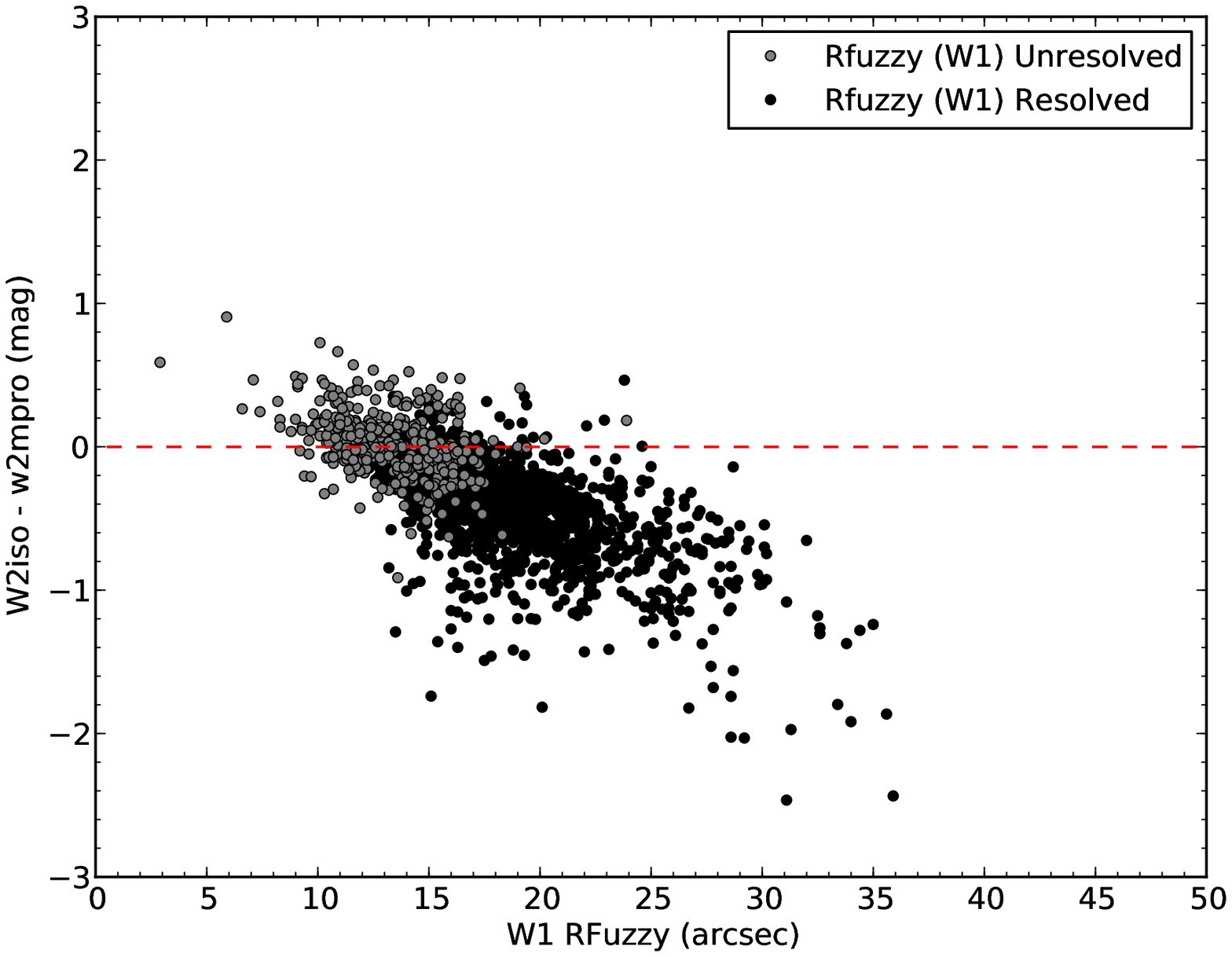}}
\vfill
\subfigure[]{\includegraphics[width=8cm]{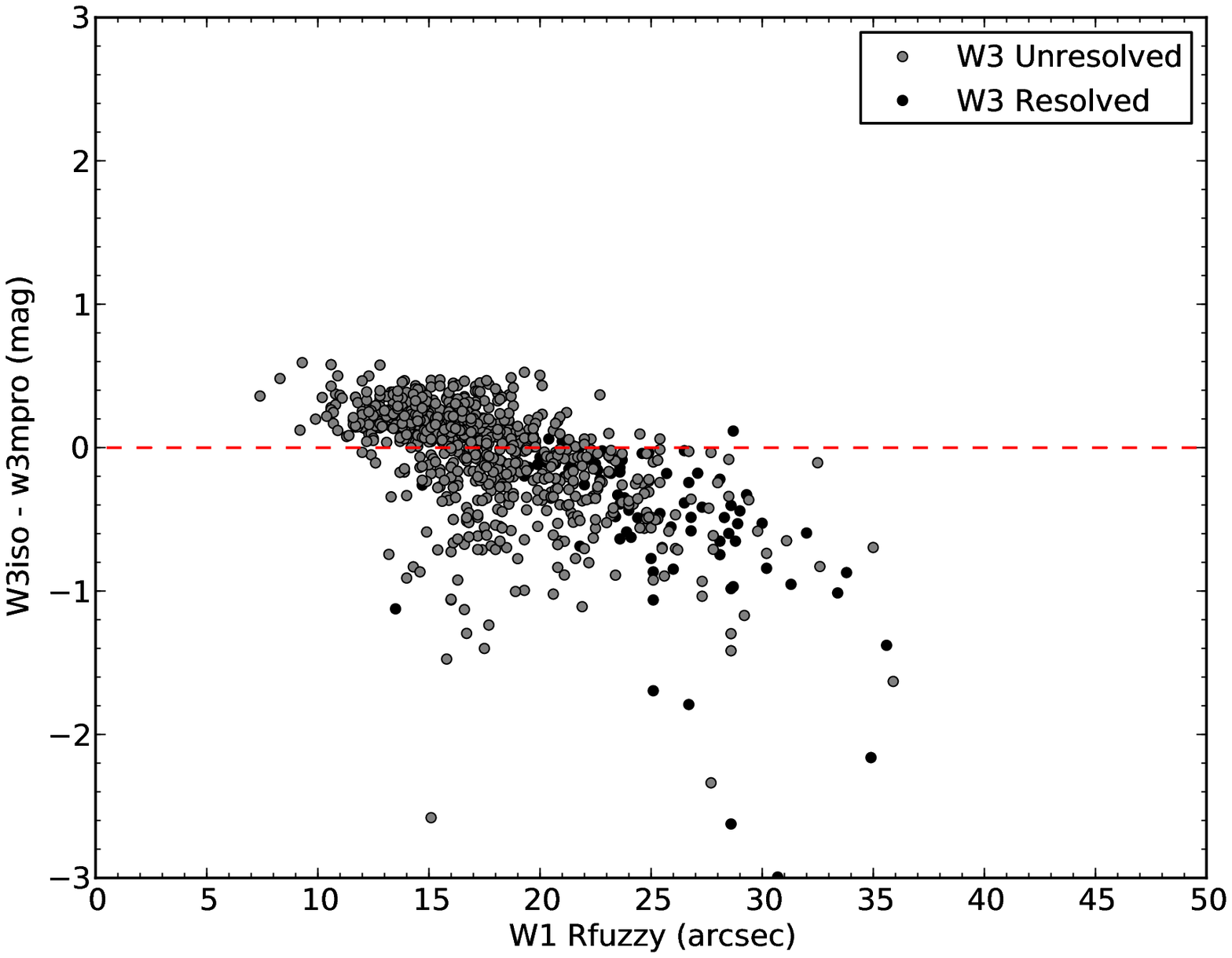}}

\caption{The difference between a 1$\sigma$ isophotal measurement (\S\,3.1.) and a profile-fit measurement, as a function of the Rfuzzy parameter (with low signal to noise sources removed) for 2590 sources in G15.}
\label{delmag_rfuzz_sn}
\end{center}
\end{figure*}

In Figure \ref{delmag_rfuzz_sn} we test 2590 sources in G15 chosen using the prescription of \S\,3, i.e. most of which are expected to be resolved in W1. This shows that the largest offset between the W1iso and {\it w1mpro} photometry measurements occurs for sources deemed resolved by the Rfuzzy parameter. This behavior is consistent with sources whose flux is underestimated by profile fitting i.e. resolved. Unresolved sources show the expected scatter around 0. In Figure \ref{delmag_rfuzz_sn}c we see that few sources are resolved in the W3 band.

\subsection{Behaviour of Resolved Photometry}\label{resphot}

\begin{figure*}[hpt]
\begin{center}
\subfigure[]{\includegraphics[width=8cm]{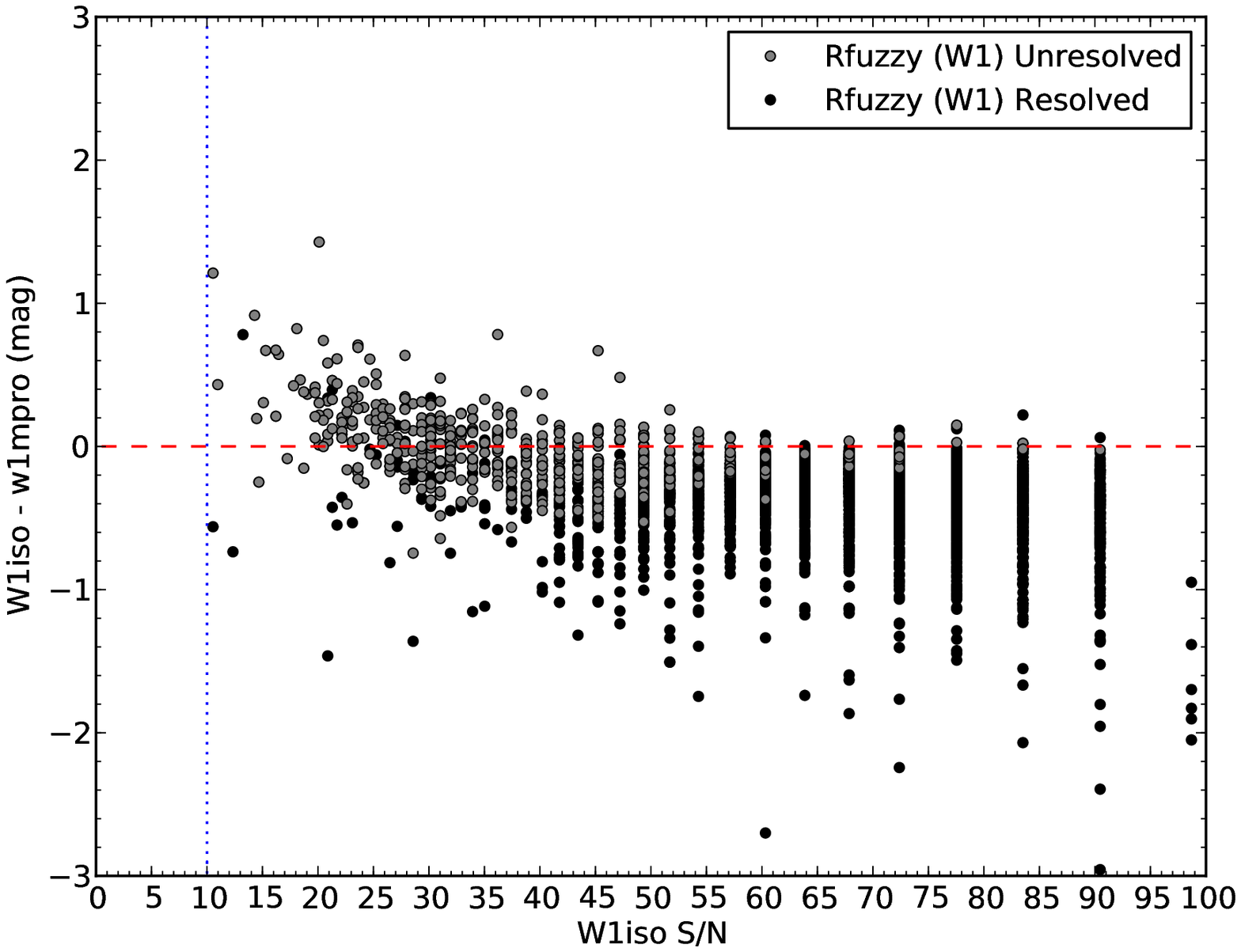}}
\hfill
\subfigure[]{\includegraphics[width=8cm]{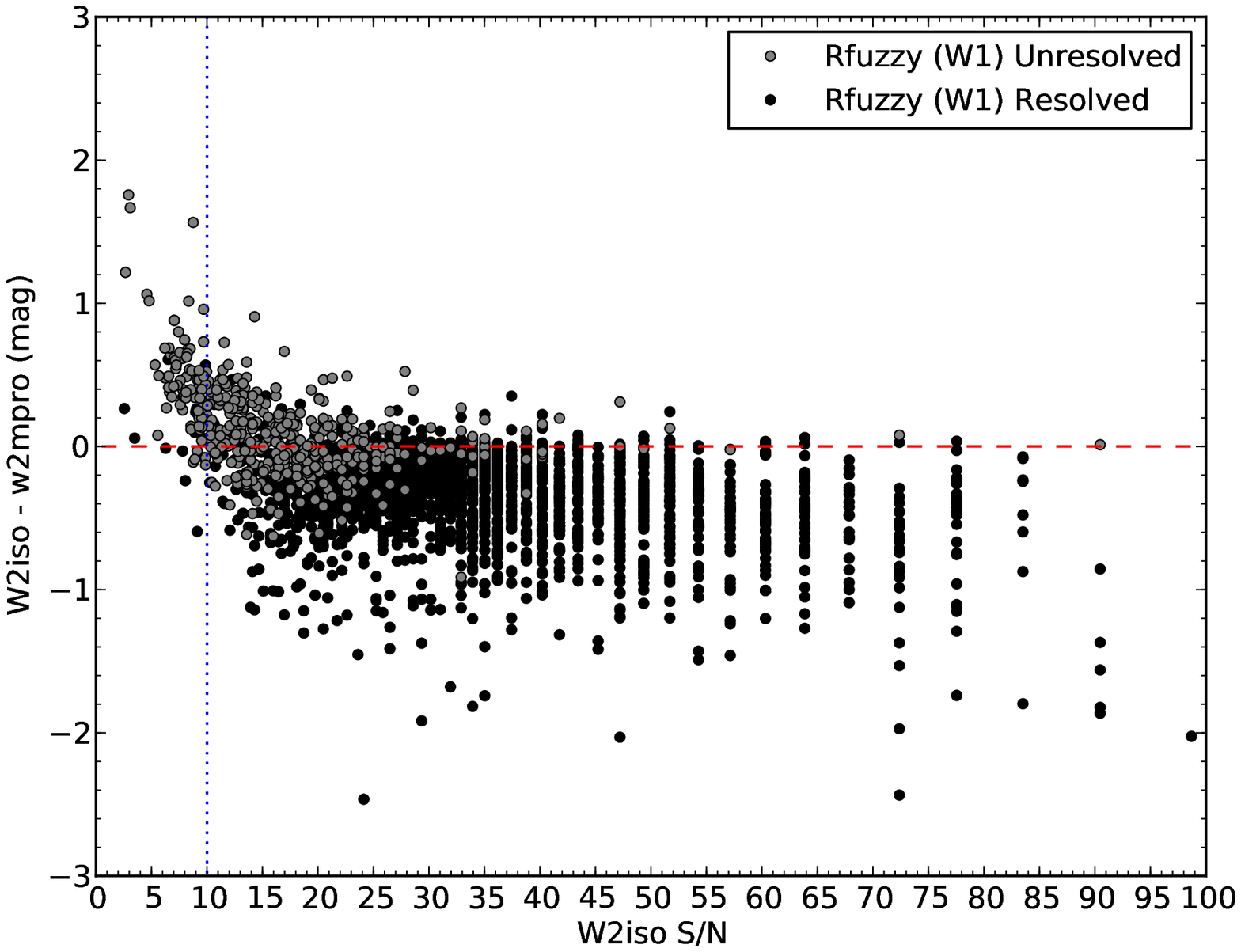}}

\subfigure[]{\includegraphics[width=8cm]{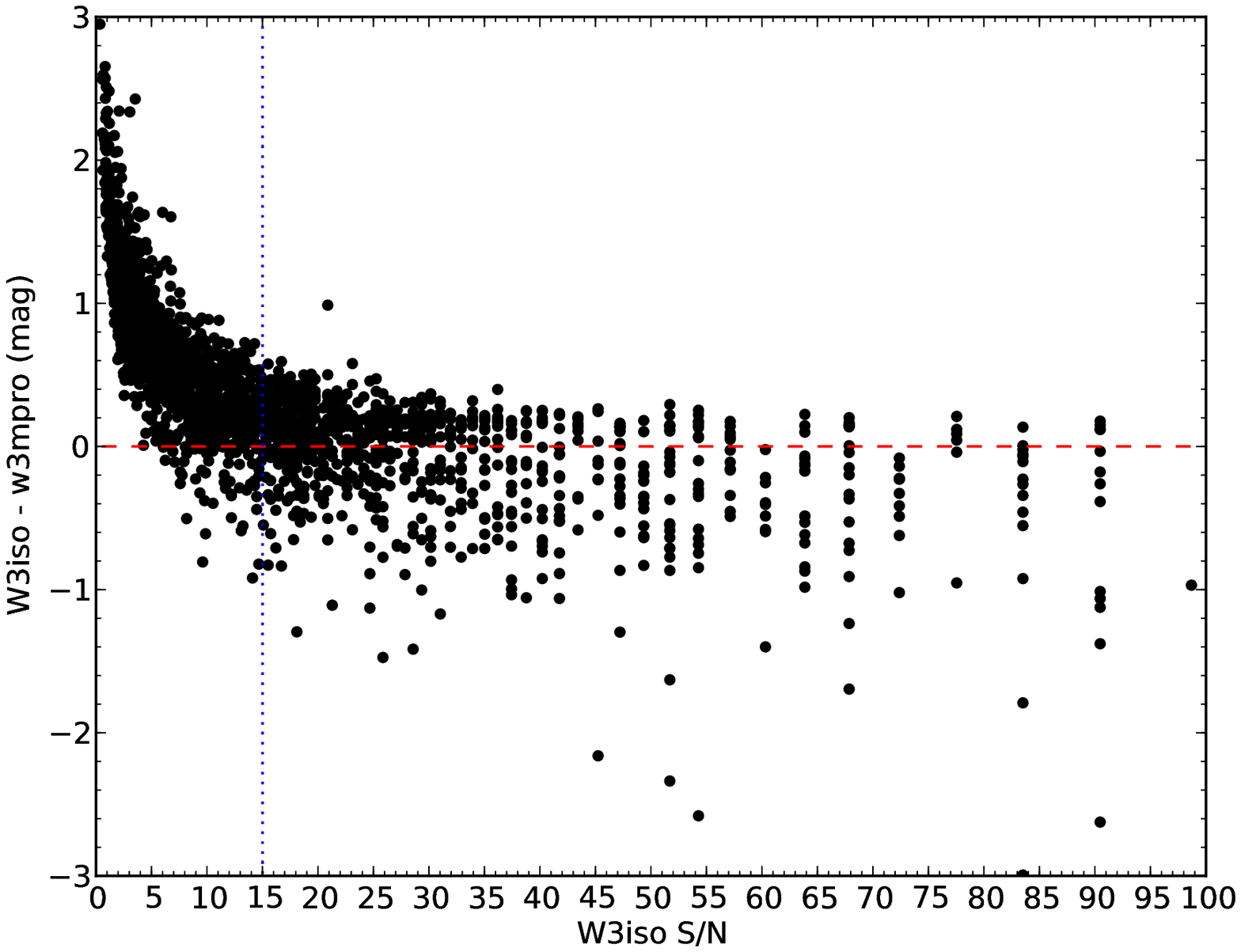}}
\hfill
\subfigure[]{\includegraphics[width=8cm]{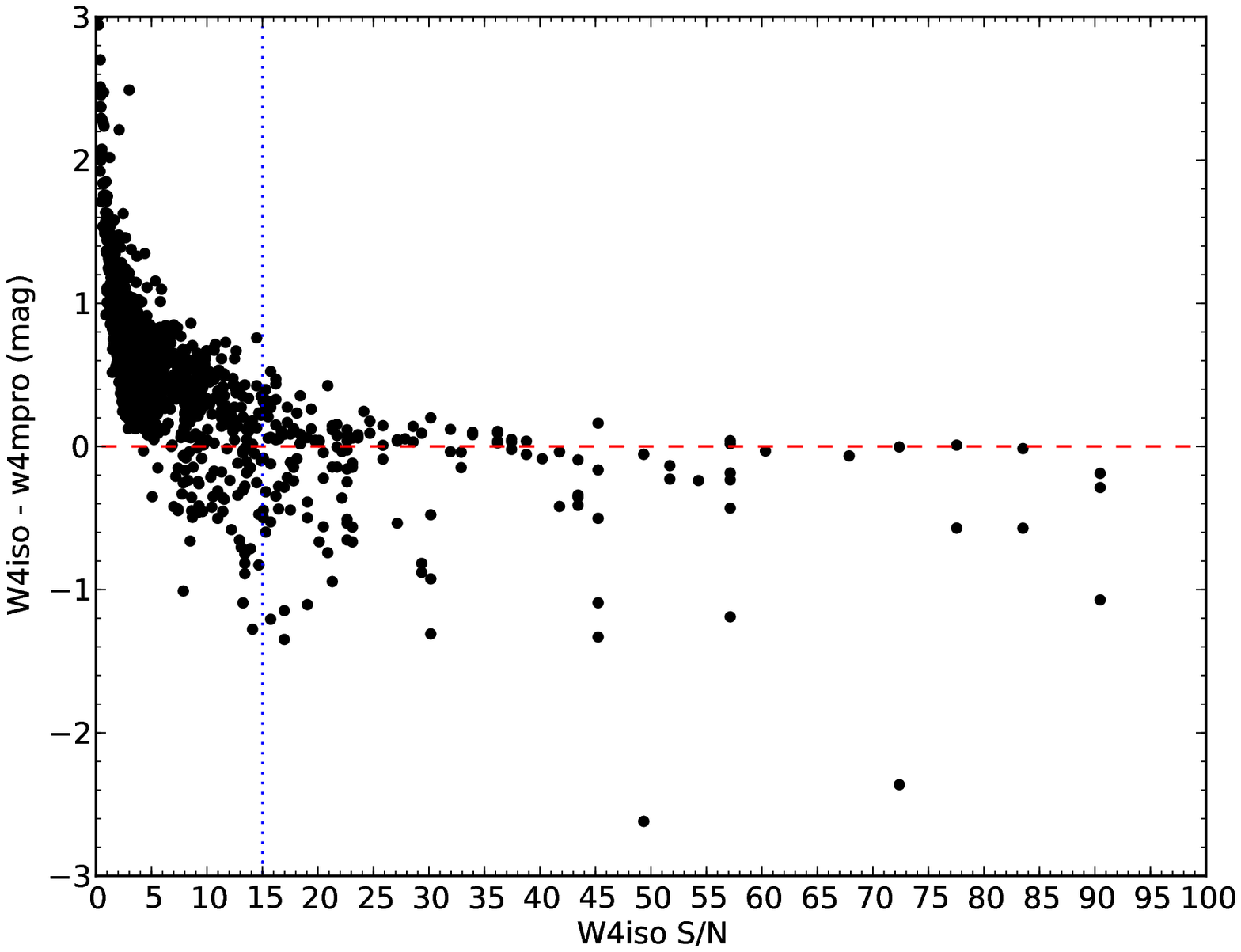}}
\caption{The difference between an isophotal measurement and a profile-fit measurement as a function of signal to noise (S/N) in the isophotal measurement in each band. The vertical dotted lines indicate the S/N limits where the isophotal photometry becomes less reliable and {\it w$\star$mpro} photometry is recommended.}
\label{delmag_snr}
\end{center}
\end{figure*}

In order to establish the reliablity of isophotal photometry, we use our test sample (2590 potentially resolved sources from the G15 GAMA field). In Figure \ref{delmag_snr} we compare the difference between W$\star$iso and {\it w$\star$mpro} photometry as a function of S/N ratio. At low S/N the isophotal measurement becomes unreliable -- in some cases the fluxes are inflated, notably for W3 and W4  --  due to contamination from background sources and {\it w$\star$mpro} provide the most robust measurement. 

\begin{figure*}
\begin{center}
\subfigure[]{\includegraphics[width=8cm]{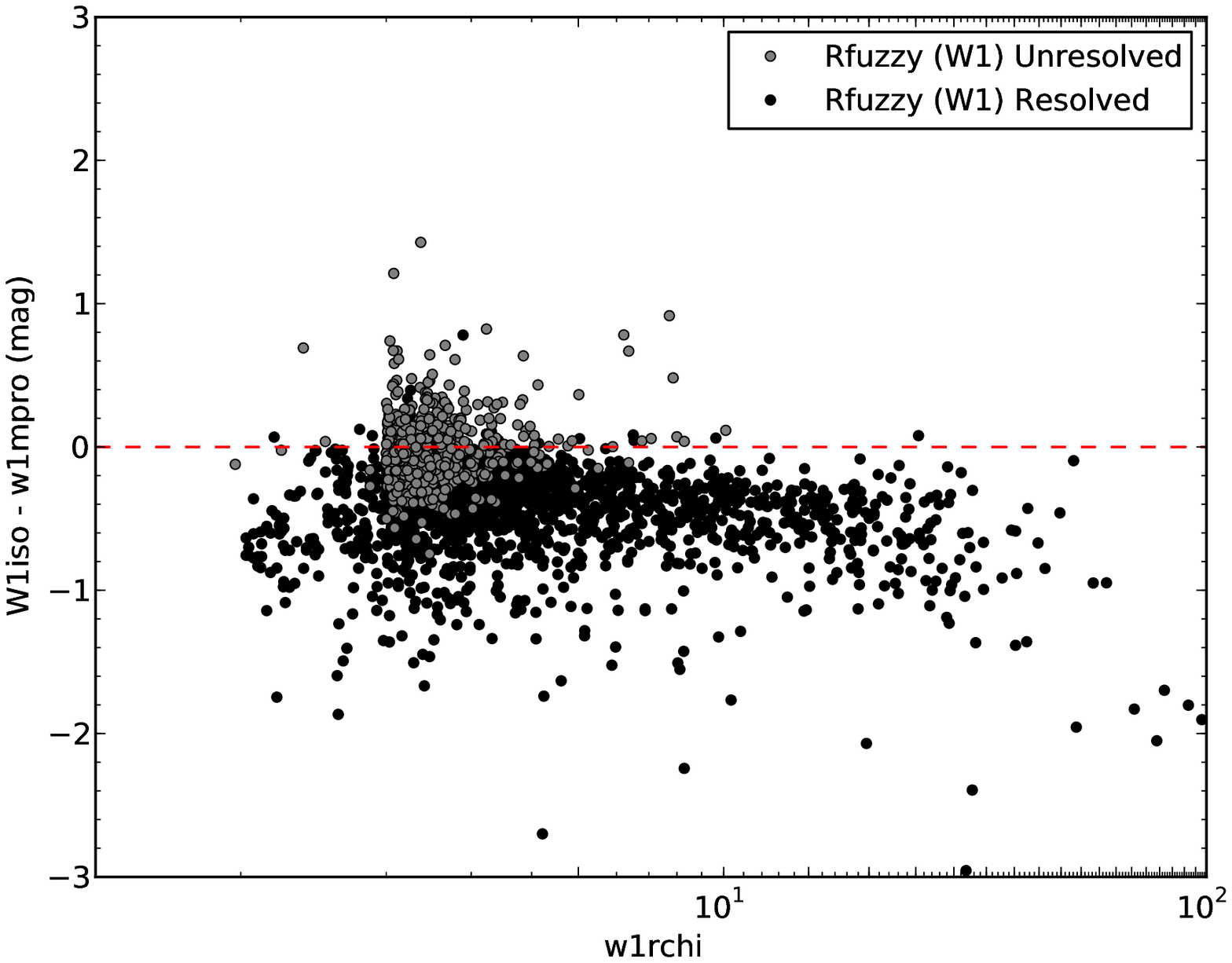}}
\hfill
\subfigure[]{\includegraphics[width=8cm]{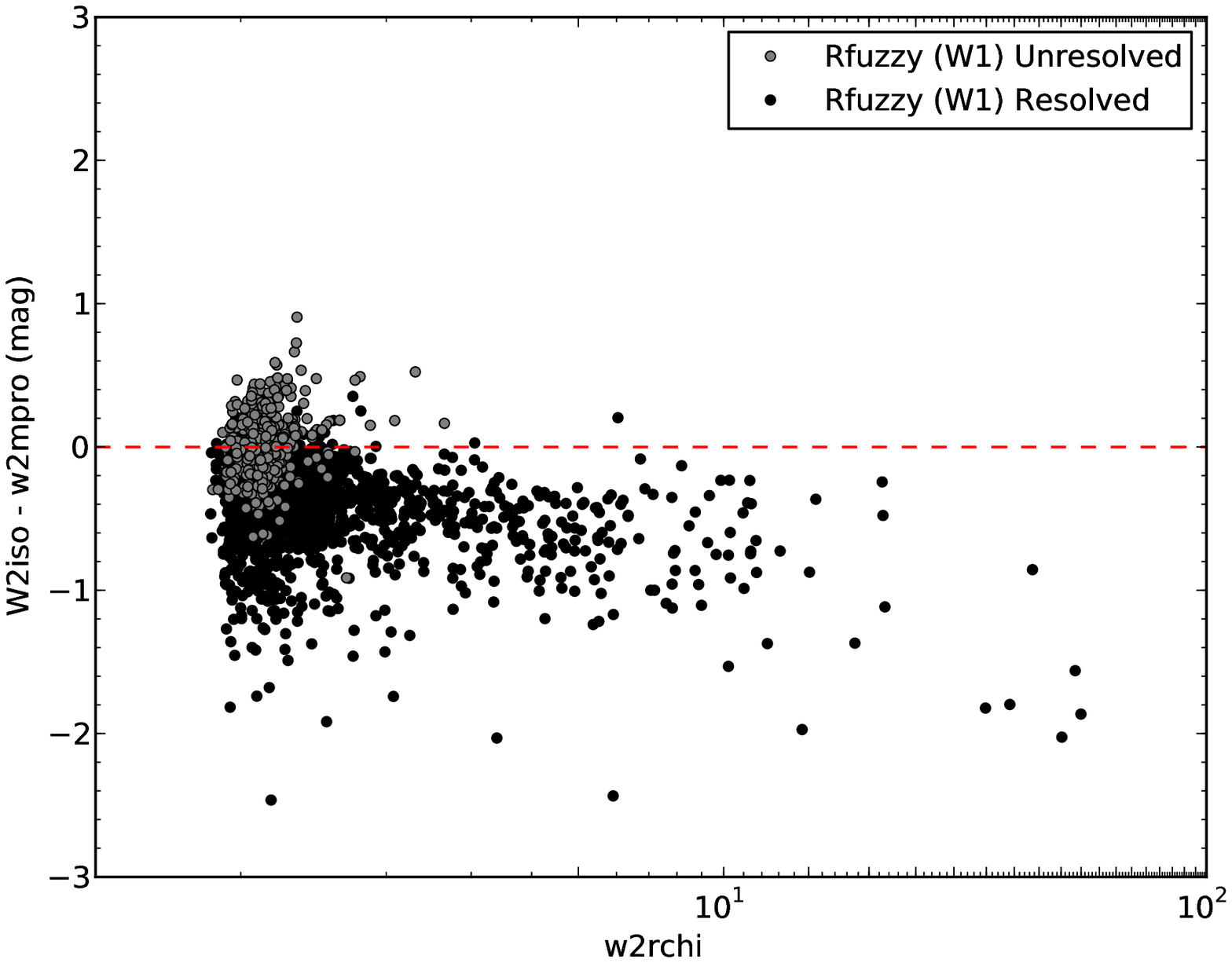}}
\vfill
\subfigure[]{\includegraphics[width=8cm]{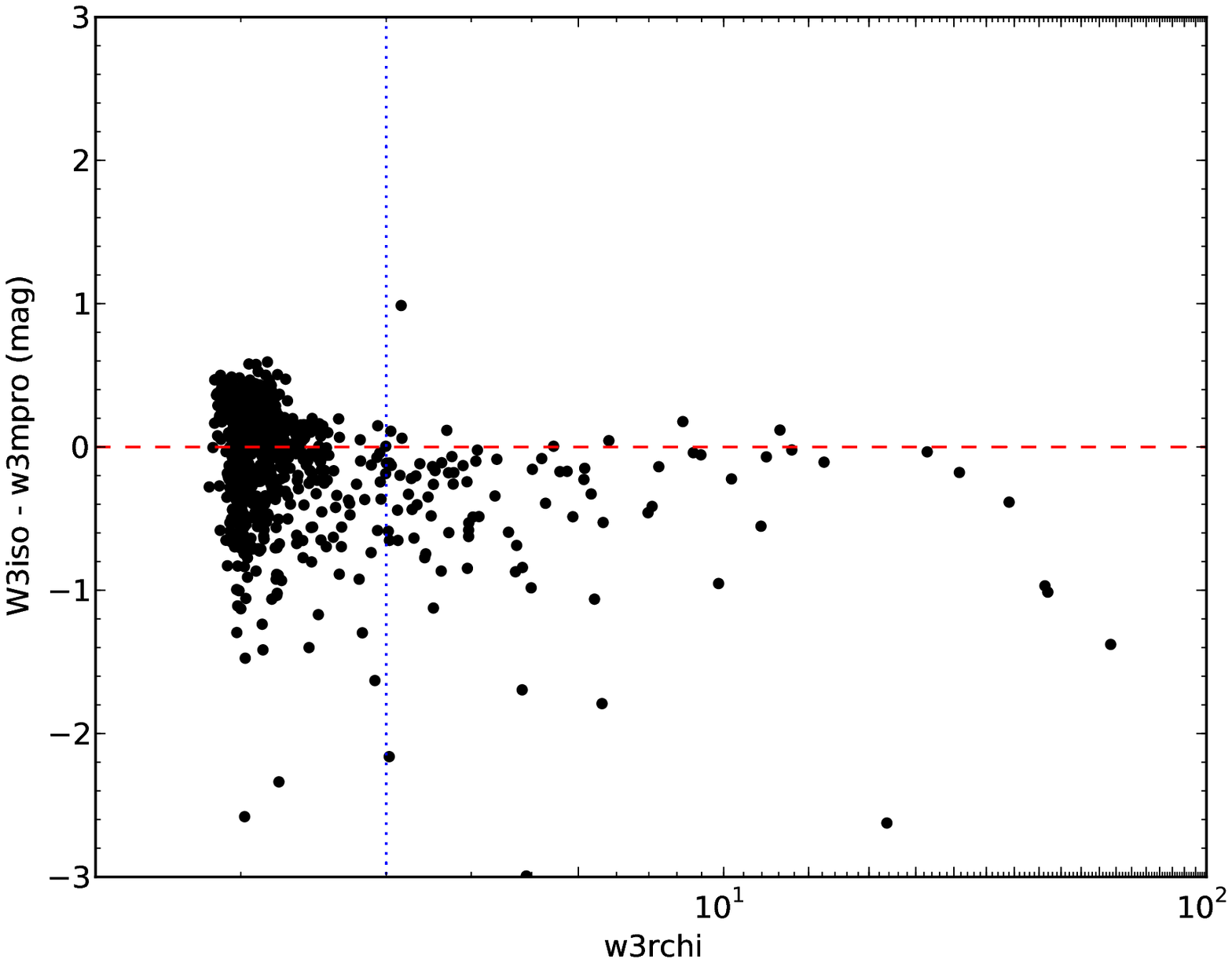}}
\hfill
\subfigure[]{\includegraphics[width=8cm]{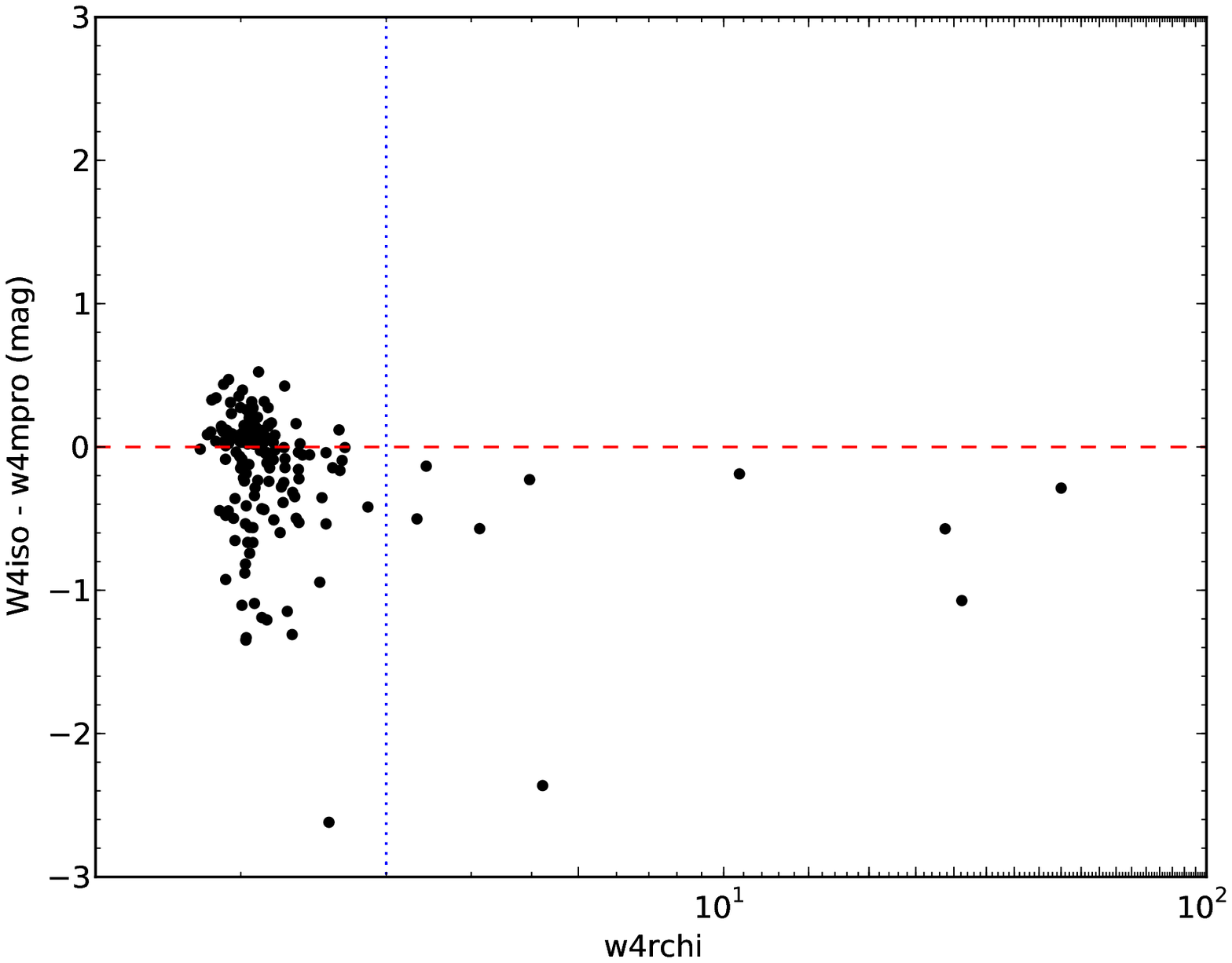}}
\caption{The difference between an isophotal measurement and a profile-fitted measurement as a function of reduced $\chi^2$ with low signal to noise sources removed. The vertical lines indicate the limits for resolved sources in the W3 and W4 bands.}
\label{delmag_rchi_sn}
\end{center}
\end{figure*}

In Figure \ref{delmag_rchi_sn} we again use the difference between W$\star$iso and {\it w$\star$mpro} photometry to illustrate that the sensitivity of the W1 and W2 bands prevents the {\it w1rchi2} and {\it w2rchi} values from acting as a reliable discriminator of resolvedness. For W3 and W4, however, it can be used and the limits derived from the plots are indicated.

\begin{figure*}
\begin{center}
\subfigure[3.4\micron]{\includegraphics[width=8cm]{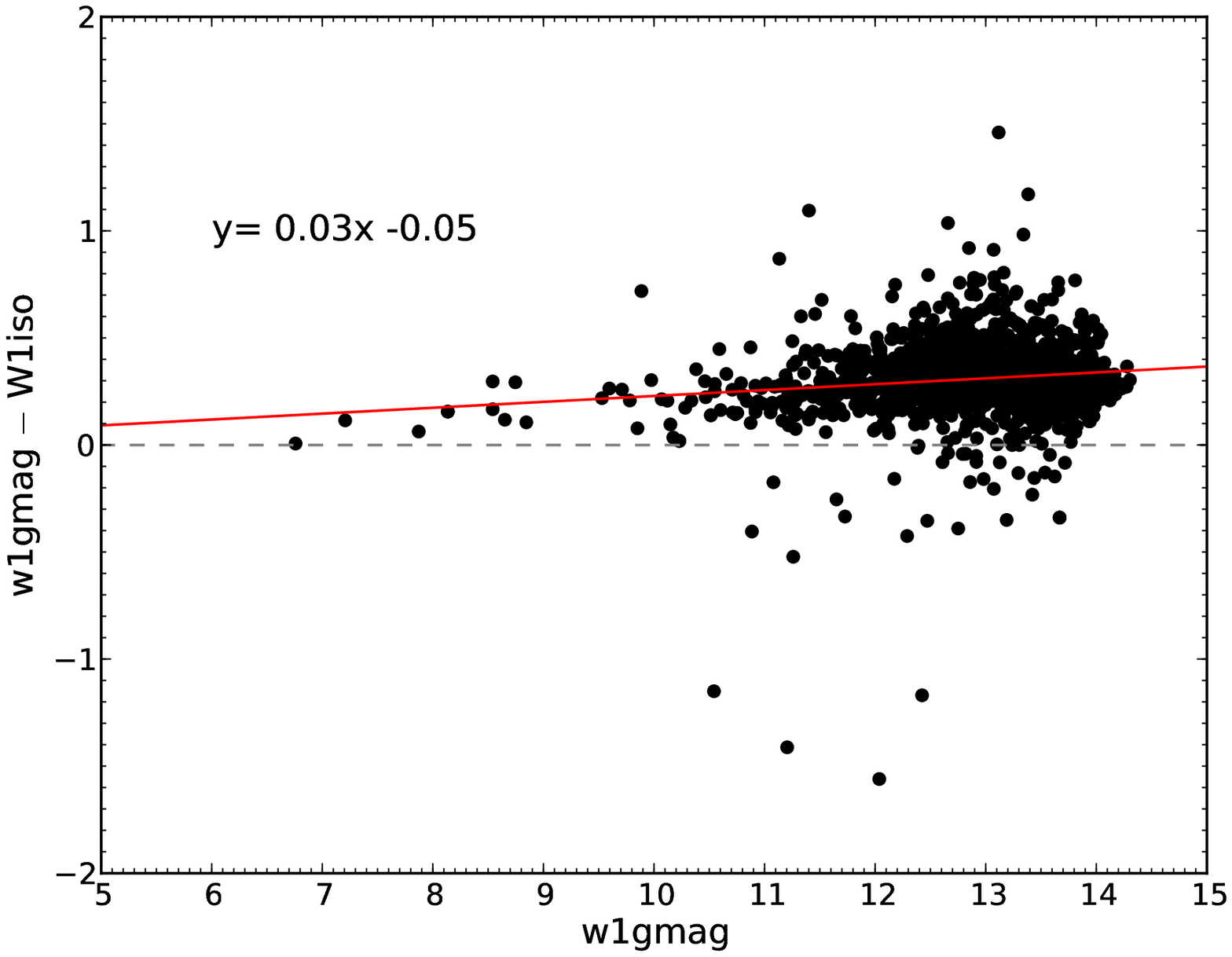}}
\hfill
\subfigure[4.6\micron]{\includegraphics[width=8cm]{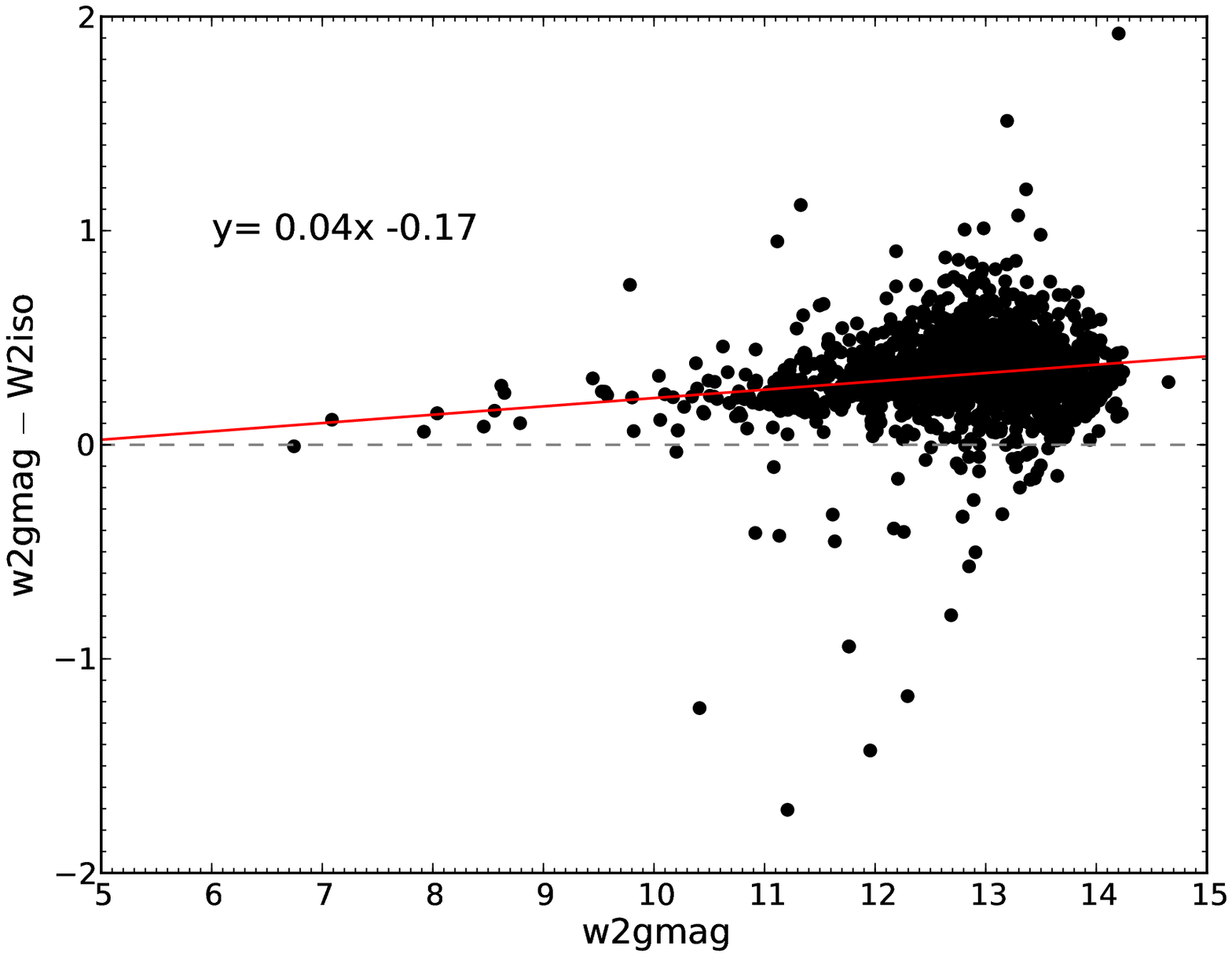}}
\caption{Comparison of {\it w$\star$gmag} and W$\star$iso photometry as a function of {\it w$\star$gmag}.}
\label{resplots}
\end{center}
\end{figure*}

We explore the relationship between the isophotal photometry and the {\it w$\star$gmag} photometry for sources in the 2MASS XSC. Since the {\it w$\star$gmag}s are measured using elliptical apertures with radii scaled to twice the 2MASS radii, this provides an indication of the additional sensitivity. We note that {\it w$\star$gmag}s can be contaminated by nearby objects since no attempt is made to remove neighboring sources. In Figure \ref{resplots} we show the difference between {\it w$\star$gmag} and W$\star$iso photometry for resolved galaxies in G12 and G15. This shows that for fainter sources {\it w$\star$gmag} is brighter than W$\star$iso, likely due to contamination.

\begin{figure*}
\begin{center}
\subfigure[3.4\micron]{\includegraphics[width=8cm]{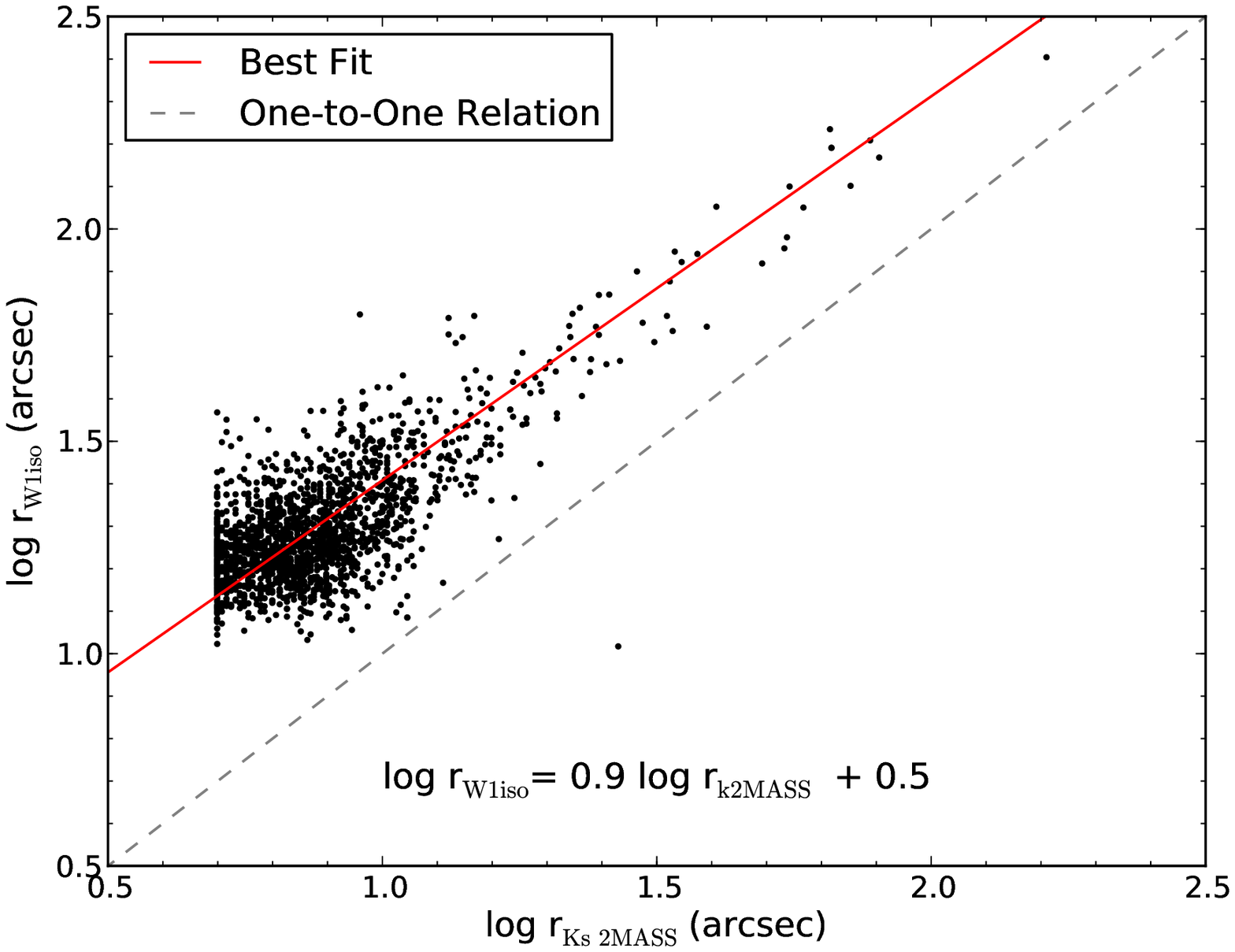}}
\hfill
\subfigure[4.6\micron]{\includegraphics[width=8cm]{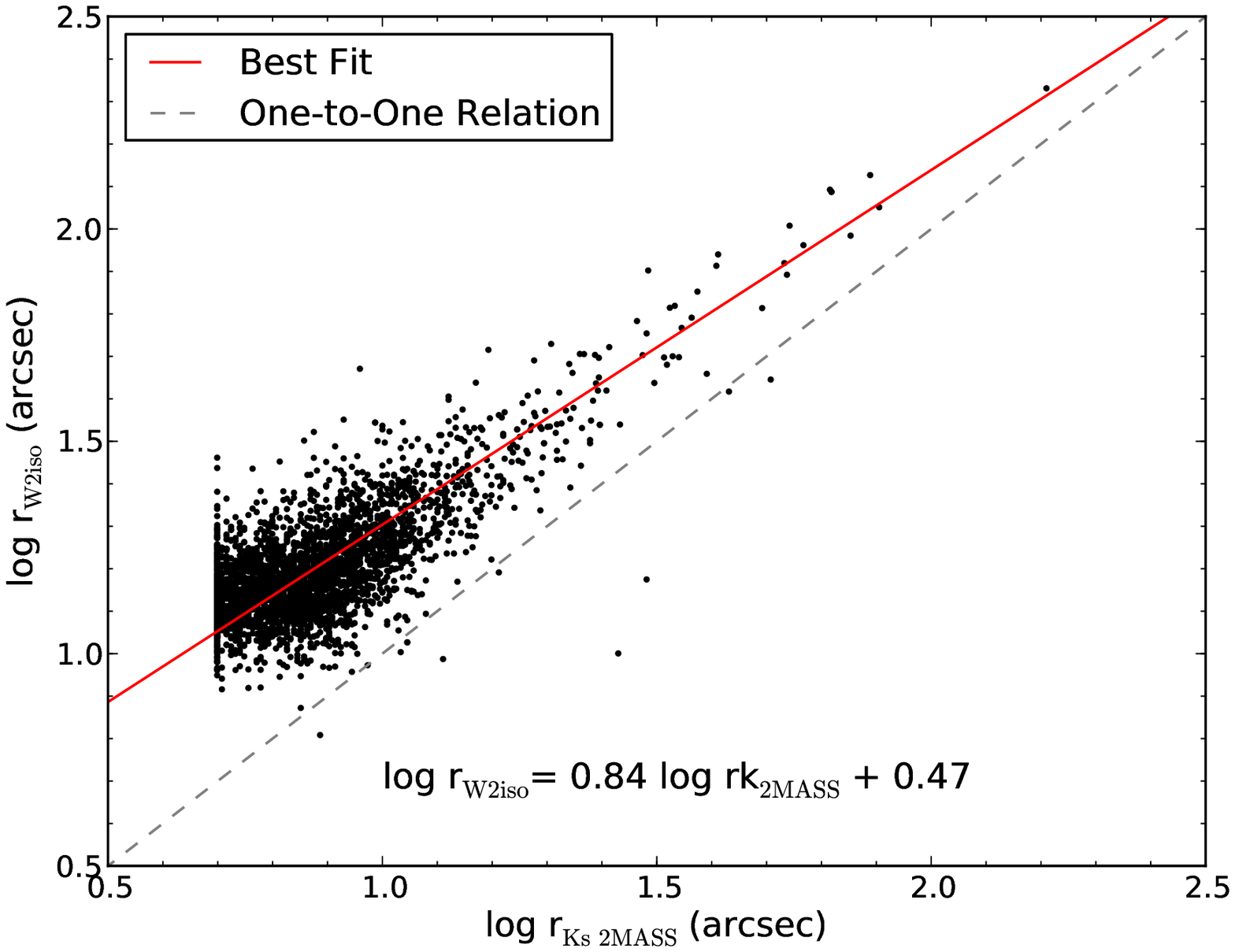}}
\caption{Comparison of isophotal and {\it w$\star$gmag} radii for resolved sources in G12 and G15; the dashed line indicates a one-to-one relation.}
\label{radplots}
\end{center}
\end{figure*}

Finally, in Figure \ref{radplots} the {\it WISE} isophotal and 2MASS isophotal radii of resolved sources which clearly shows that the 1$\sigma$ isophotal radii are systematically larger than the 2MASS radii by a factor of 2 to 2.5. The largest offset occurs for the most compact 2MASS sources probably due to the increased sensitivity of WISE in the W1 and W2 bands resolving more of the galaxy relative to 2MASS.

\end{document}